\documentclass[aps,twocolumn,showpacs,superscriptaddress]{revtex4}

\usepackage{graphicx}
\usepackage{bm}
\usepackage{amsmath}
\usepackage{dcolumn}%

\newcommand{\eg}{\ensuremath{e_{g}}}
\newcommand{\tg}{\ensuremath{t_{2g}}}

\newcommand{\mb}{\ensuremath{\mu_{\text{B}}}}

\newcolumntype{/}{D{/}{/}{2,2}}  
\newcolumntype{.}{D{.}{.}{0}}  

\begin{document}

\title{Resonant inelastic x-ray scattering of the $J_{eff}$ = 1/2 Mott
  insulator Sr$_2$IrO$_4$ from the density-functional theory }

\author{V.N. Antonov}

\affiliation{G. V. Kurdyumov Institute for Metal Physics of the
  N.A.S. of Ukraine, 36 Academician Vernadsky Boulevard, UA-03142
  Kyiv, Ukraine}

\affiliation{Max-Planck-Institute for Solid State Research,
  Heisenbergstrasse 1, 70569 Stuttgart, Germany}

\author{D.A. Kukusta}

\affiliation{G. V. Kurdyumov Institute for Metal Physics of the
  N.A.S. of Ukraine, 36 Academician Vernadsky Boulevard, UA-03142
  Kyiv, Ukraine}

\author{L.V. Bekenov}

\affiliation{G. V. Kurdyumov Institute for Metal Physics of the
  N.A.S. of Ukraine, 36 Academician Vernadsky Boulevard, UA-03142
  Kyiv, Ukraine}

\date{\today}

\begin{abstract}

We have investigated the electronic structure of Sr$_2$IrO$_4$ within the
density-functional theory (DFT) using the generalized gradient approximation
while taking into account strong Coulomb correlations (GGA+$U$) in the
framework of the fully relativistic spin-polarized Dirac linear muffin-tin
orbital band-structure method. We have investigated the x-ray absorption
spectra (XAS), x-ray magnetic circular dichroism (XMCD), and resonant
inelastic x-ray scattering (RIXS) spectra at the Ir $L_3$ and O $K$ edges. The
calculated results are in good agreement with the experimental data. The
RIXS spectrum of Sr$_2$IrO$_4$ at the Ir $L_3$ edge in
addition to the elastic scattering peak at 0 eV possesses a sharp feature
below 1.5 eV corresponding to transitions within the Ir {\tg} levels. The
excitation located from 2 eV to 5 eV is due to {\tg} $\rightarrow$ {\eg}
transitions. The third wide structure situated at 5$-$12 eV appears due to
transitions between the Ir 5$d_{\rm{O}}$ states derived from the "tails" of
oxygen 2$p$ states and {\eg} and {\tg} states. The RIXS spectrum of
Sr$_2$IrO$_4$ at the O $K$ edge consists of three major inelastic excitations
at 0.7 eV, 3.5 eV, and around 6.2 eV. We have found that the first low energy
feature is due to interband transitions between occupied and empty O$_{\tg}$
transitions, which appear due to the strong hybridization between oxygen 2$p$
and Ir {\tg} states in the close vicinity of the Fermi level. The next two peaks at
around 3.5 and 6.2 eV reflect the iterband transitions from the occupied O
2$p$ states and the empty oxygen states which arise from the
hybridization with Ir {\tg} and {\eg} states, respectively. We have found that the
theory reproduces well the shape and energy position of the low energy feature,
but the second and the third peaks are shifted towards smaller energy in
comparison with the experimental measurements. To reproduce the correct energy
position of the oxygen 2$p$ band we have used a self-interaction-like correction
procedure. We have found that the dependence of the RIXS spectrum at the oxygen
$K$ edge on the incident photon energy and the momentum transfer vector {\bf
  Q} is much stronger in comparison with the correspondent dependence at
the Ir $L_3$ edge.

\end{abstract}


\maketitle

\section{Introduction}

\label{sec:introd}

In 5$d$ transition metal compounds the energy scale of the spin-orbit coupling
(SOC) is comparable to the on-site Coulomb interaction and the crystal-field
energy. Due to the strong competition between these interactions 
fascinating electronic states can arise. The SOC in such systems splits the
{\tg} orbitals into a quartet ($j_{eff}$ = 3/2) and a doublet
($j_{eff}$ = 1/2) \cite{JaKh09,CPB10,WCK+14}. In 5$d^5$ (Ir$^{4+}$) iridium
oxides, the quartet $j_{eff}$ = 3/2 is fully occupied, and the relatively narrow
$j_{eff}$ = 1/2 doublet occupied by one electron can be splitted by a moderate
Hubbard $U_{eff}$ with opening a small band gap called the relativistic Mott
gap \cite{KJM+08,MAV+11,AUU18}. Iridates have been at the center of an
intensive search in recent years for novel phenomena, such as topological
insulators \cite{QZ10,Ando13,WBB14,BLD16}, Mott insulators
\cite{KJM+08,KOK+09,JaKh09,WSY10,MAV+11}, Weyl semimetals
\cite{WiKi12,GWJ12,SHJ+15}, and quantum spin liquids \cite{JaKh09,KAV14}.

Among iridium compounds, Sr$_2$IrO$_4$, a single-layer member of the
Ruddlesden-Popper series iridates, is of special interest. It has a
quasi-two-dimensional (2D) square-lattice perovskite structure and was the
first discovered spin-orbit $j_{eff}$ = 1/2 Mott insulator
\cite{KJM+08}. Besides, it has the structural and physical similarity with
La$_2$CuO$_4$, a parent compound to high-$T_c$ cuprates, such as the presence
of a pseudogap \cite{KJM+15,YRX+15,BBF+17}, similar Fermi surfaces and Fermi
arcs (in electron- and hole-doped compounds) \cite{DHS03,KKD+14}, $d$-wave
symmetry \cite{KSD+16,TSF+17}, electron-boson coupling \cite{HCP+19}, and
similarities in the magnetic ordering and magnetic excitations
\cite{KCU+12,YCY+19}. However, the superconductivity in Sr$_2$IrO$_4$ has not
been found yet \cite{BKK19}.

In this work we focus our attention on the RIXS properties in
Sr$_2$IrO$_4$. Since the first publication by Kao {\it et al.} on NiO
\cite{KCH+96}, the resonant inelastic X-ray scattering method has shown
remarkable progress in condensed matter physics research as a spectroscopic
technique to record the momentum and energy dependence of inelastically
scattered photons in complex materials \cite{AVD+11}. RIXS nowadays has rapidly
become the forefront of experimental photon science. It provides a
direct probe of spin and orbital states and dynamics. RIXS has a number of
unique features in comparison with other spectroscopic techniques. It covers a
large scattering phase space and requires only small sample volumes. It also
is bulk sensitive, polarization dependent, as well as element and orbital
specific \cite{AVD+11}.

Depending on the x-ray resonant energy, RIXS can be divided into two classes:
soft x-ray and hard x-ray \cite{AVD+11}. For high atomic number transition
metal elements, such as 5$d$ transition metal compounds, the $L$-edge resonant
energies are in the hard x-ray region. For such spectra high quality single
crystals are needed as the key optical elements. The RIXS resolution crucially
depends on the availability of a high quality single crystal with a Bragg
diffraction peak close to back-scattering at the energy of the $L$-edge of the
targeted element. This requirement severely limits the application of RIXS,
and thus by far the majority of hard x-ray RIXS studies have been focused on
5$d$ iridates \cite{LKH+12,HGC+14,CGL+16,NBC+18,TKG+19,ACC+22} and osmates
\cite{CVB+16,TCM+17,CVB+17}. In a soft RIXS setup, the x-ray energy range is
usually below about 2 keV \cite{LDL+15}. The $L$-edges of the 3$d$ transition
metal elements all fall below this energy scale. The energy resolution in the soft
x-ray region is relatively high. For example, the combined energy resolution
was 150 meV at the Ni $L_3$ edge ($\sim$850 eV) in Ta$_2$NiSe$_5$
\cite{MHP+20}. The best energy resolution currently achieved for RIXS at the
oxygen $K$-edge ($\sim$530 eV) of the common ligand atoms is about 45-50 meV
\cite{LJM+13,LOH+18}, which is much better than the majority of 5$d$ elements
probed using $L$-edge RIXS in the hard x-ray region up to date. We should
mention, however, that in recent years the experimentalists have achieved a
remarkable progress in increasing the resolution for hard RIXS spectra. For
example, Kim {\it et al.} \cite{KDK+23} have been able to obtain the total
resolution of 34.2 meV at the Ir $L_3$ edge in Sr$_2$IrO$_4$. Such
resolution permits direct measurements of single-magnon excitations as well
as other many-body excitations in strongly correlated systems.

In the x-ray absorption (XA), x-ray magnetic circular dichroism
(XMCD), and RIXS processes at the O $K$ edge, the 1$s$ core level is
involved. The exchange splitting of the 1$s$-core state is extremely
small and SOC is absent in the O 1$s$ orbitals, therefore, only the
exchange and spin-orbit splitting of the 2$p$ states is responsible
for the observed spectra at the oxygen $K$ edge. On the other hand,
the oxygen valence 2$p$-states of the surrounding ligand atoms are
sensitive to the electronic states at neighboring sites because of
their delocalized nature. They strongly hybridize with the 5$d$
orbitals. Due to such hybridization combined with high SOC at the 5$d$
ion, information on the elementary excitations can be extracted using
an indirect RIXS process at the O $K$ edge \cite{LOH+18}. Although O
$K$ RIXS has a much smaller penetration depth ($\sim$100 nm) than 5$d$
$L$ RIXS, a comparison between O $K$ and Ir $L_3$ spectra measured on
Sr$_2$IrO$_4$ suggests that they have a comparable counting efficiency
\cite{LOH+18}. The lower penetration depth of soft x-rays has its own
advantages providing high sensitivity to ultrathin samples such as
films. Soft x-ray RIXS at the O $K$ edge is a promising method for
studying the electronic and magnetic excitations in 5$d$
compounds. There are several experimental investigations of the RIXS
spectra at the oxygen $K$ edge \cite{LDL+15,LOH+18,PTP+20,KDK+23} in
Sr$_2$IrO$_4$. The Ir $L_3$ RIXS spectra in this oxide are
investigated in
Refs. \cite{IJY+11,KCU+12,LCG+14,PTP+20,BKP+20,KDK+23,CGL+23}.

We carry out here a detailed study of the electronic structure, the
XA, XMCD, and RIXS spectra of Sr$_2$IrO$_4$ in terms of the density
functional theory. Our study sheds light on the important role of band
structure effects and transition metal 5$d$ $-$ oxygen 2$p$
hybridization in the spectral properties in 5$d$ oxides. The energy
band structure, the XA, XMCD, and RIXS spectra of Sr$_2$IrO$_4$ are
investigated in the {\it ab initio} approach using the fully
relativistic spin-polarized Dirac linear muffin-tin orbital
band-structure method. We use both the generalized gradient
approximation (GGA) and the GGA+$U$ approach to assess the sensitivity
of the RIXS results to different treatment of the correlated
electrons.

The paper is organized as follows. The crystal structure of
Sr$_2$IrO$_4$ and computational details are presented in
Sec. II. Sec. III presents the electronic and magnetic structures of
Sr$_2$IrO$_4$. In Sec. IV the theoretical investigations of the XA,
XMCD, and RIXS spectra of Sr$_2$IrO$_4$ at the Ir $L_{2,3}$ edges are
presented, the theoretical results are compared with the experimental
measurements. In Sec. V we present the theoretical investigations of
the XA and RIXS spectra at the O $K$ edge. Finally, the results are
summarized in Sec. VI.

\section{Computational details}
\label{sec:details}

\paragraph{X-ray magnetic circular dichroism.} 

Magneto-optical (MO) effects refer to various changes in the
polarization state of light upon interaction with materials possessing
a net magnetic moment, including rotation of the plane of linearly
polarized light (Faraday, Kerr rotation), and the complementary
differential absorption of left and right circularly polarized light
(circular dichroism). In the near visible spectral range these effects
result from excitation of electrons in the conduction band. Near x-ray
absorption edges, or resonances, magneto-optical effects can be
enhanced by transitions from well-defined atomic core levels to
transition symmetry selected valence states.

Within the one-particle approximation, the absorption coefficient
$\mu^{\lambda}_j (\omega)$ for incident x-ray polarization $\lambda$ and
photon energy $\hbar \omega$ can be determined as the probability of
electronic transitions from initial core states with the total angular
momentum $j$ to final unoccupied Bloch states

\begin{eqnarray}
\mu_j^{\lambda} (\omega) &=& \sum_{m_j} \sum_{n \bf k} | \langle \Psi_{n \bf k} |
\Pi _{\lambda} | \Psi_{jm_j} \rangle |^2 \delta (E _{n \bf k} - E_{jm_j} -
\hbar \omega ) \nonumber \\
&&\times \theta (E _{n \bf k} - E_{F} ) \, ,
\label{mu}
\end{eqnarray}
where $\Psi _{jm_j}$ and $E _{jm_j}$ are the wave function and the
energy of a core state with the projection of the total angular
momentum $m_j$; $\Psi_{n\bf k}$ and $E _{n \bf k}$ are the wave
function and the energy of a valence state in the $n$-th band with the
wave vector {\bf k}; $E_{F}$ is the Fermi energy.

$\Pi _{\lambda}$ is the electron-photon interaction
operator in the dipole approximation
\begin{equation}
\Pi _{\lambda} = -e \mbox{\boldmath$\alpha $} \bf {a_{\lambda}},
\label{Pi}
\end{equation}
where $\bm{\alpha}$ are the Dirac matrices and $\bf {a_{\lambda}}$ is the
$\lambda$ polarization unit vector of the photon vector potential,
with $a_{\pm} = 1/\sqrt{2} (1, \pm i, 0)$,
$a_{\parallel}=(0,0,1)$. Here, $+$ and $-$ denote, respectively, left
and right circular photon polarizations with respect to the
magnetization direction in the solid. Then, x-ray magnetic circular
and linear dichroisms are given by $\mu_{+}-\mu_{-}$ and
$\mu_{\parallel}-(\mu_{+}+\mu_{-})/2$, respectively.  More detailed
expressions of the matrix elements in the electric dipole
approximation may be found in
Refs.~\cite{GET+94,book:AHY04,AHS+04}.  The matrix elements due
to magnetic dipole and electric quadrupole corrections are presented
in Ref.~\cite{AHS+04}.

\paragraph{Resonant inelastic x-ray scattering.} 

In the direct RIXS process \cite{AVD+11} the incoming photon with
energy $\hbar \omega_{\mathbf{k}}$, momentum $\hbar \mathbf{k}$, and
polarization $\bm{\epsilon}$ excites the solid from the ground state
$|{\rm g}\rangle$ with energy $E_{\rm g}$ to the intermediate state
$|{\rm I}\rangle$ with energy $E_{\rm I}$. During relaxation an
outgoing photon with energy $\hbar \omega_{\mathbf{k}'}$, momentum
$\hbar \mathbf{k}'$ and polarization $\bm{\epsilon}'$ is emitted, and
the solid is in state $|f \rangle$ with energy $E_{\rm f}$. As a
result, an excitation with energy $\hbar \omega = \hbar
\omega_{\mathbf{k}} - \hbar \omega_{\mathbf{k}'}$ and momentum $\hbar
\mathbf{q}$ = $\hbar \mathbf{k} - \hbar \mathbf{k}'$ is created.  Our
implementation of the code for the calculation of the RIXS intensity
uses Dirac four-component basis functions \cite{NKA+83} in the
perturbative approach \cite{ASG97,ASG97E}. RIXS is the second-order
process, and its intensity is given by

\begin{eqnarray}
I(\omega, \mathbf{k}, \mathbf{k}', \bm{\epsilon}, \bm{\epsilon}')
&\propto&\sum_{\rm f}\left| \sum_{\rm I}{\langle{\rm
    f}|\hat{H}'_{\mathbf{k}'\bm{\epsilon}'}|{\rm I}\rangle \langle{\rm
    I}|\hat{H}'_{\mathbf{k}\bm{\epsilon}}|{\rm g}\rangle\over
  E_{\rm g}-E_{\rm I}} \right|^2 \nonumber \\ && \times
\delta(E_{\rm f}-E_{\rm g}-\hbar\omega),
\label{I1}
\end{eqnarray}
where the RIXS perturbation operator in the dipole approximation is
given by the lattice sum $\hat{H}'_{\mathbf{k}\bm{\epsilon}}=
\sum_\mathbf{R}\hat{\bm{\alpha}}\bm{\epsilon} \exp(-{\rm
  i}\mathbf{k}\mathbf{R})$, where $\bm{\alpha}$ are the Dirac matrices. The
sum over the intermediate states $|{\rm I}\rangle$ includes the
contributions from different spin-split core states at the given
absorption edge. The matrix elements of the RIXS process in the frame of
fully relativistic Dirac LMTO method were presented in Ref. \cite{AKB22a}.

\paragraph{Crystal structure.} 

\begin{figure}[tbp!]
\begin{center}
\includegraphics[width=0.70\columnwidth]{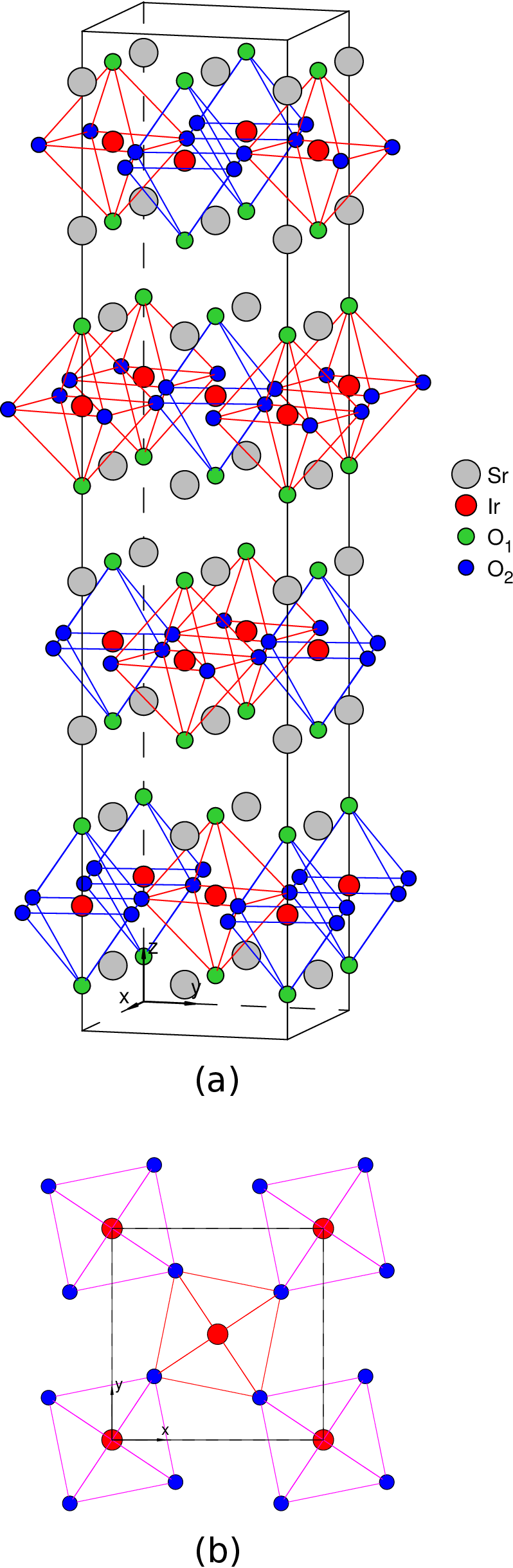}
\end{center}
\caption{\label{struc_SIO}(Color online) (a) The schematic representation
  of the body centered tetragonal $I4_1/acd$ (group number 142)
  Sr$_2$IrO$_4$ crystal structure \cite{SIN+95}; (b) the positions of
  ions in the IrO$_2$ plane perpendicular to the $c$ axis. }
\end{figure}

The powder-neutron-diffraction measurements show that Sr$_2$IrO$_4$ possesses
the tetragonal $I4_1/acd$ perovskite structure (group number 142)
[Fig. \ref{struc_SIO}(a)] \cite{SIN+95}. The IrO$_6$ octahedra in Sr$_2$IrO$_4$
are rigidly aligned, just as the CuO$_6$ octahedra in cuprates, rotated by 
$\sim$11$^{\circ}$ about the $c$ axis in the $a-b$ plane [Fig. \ref{struc_SIO}(b)], 
and have a local distortion of 4.5\% axial elongation.

Atomic positions of Sr$_2$IrO$_4$ at 10 K (the lattice constants $a$ = 5.48164
\AA, $c$ = 25.80019 \AA) for Sr, Ir, O$_1$, and O$_2$ are (0, 1/4,
$z_{\rm{Sr}}$), (0, 1/4, 3/8), (0, 1/4, $z_{\rm{O}}$), and ($x$, $x$ + 1/4,
$z_{\rm{O}}$), respectively, with $x$ = 0.1996, $z_{\rm{Sr}}$ = 0.5506, and
$z_{\rm{O}}$ = 0.4548 \cite{SIN+95}. The oxygen atoms surrounding the Ir sites
provide an octahedral environment. The Ir$-$O$_1$ and Ir$-$O$_2$ interatomic
distances are equal to 2.05886 \AA\, and 1.97704 \AA, respectively. Around
each Ir atom there are eight Sr atoms with the Ir-Sr distance $d_{\rm{Ir-Sr}}$ =
3.34615 \AA. The Ir-Ir distance $d_{\rm{Ir-Sr}}$ = 3.87610 \AA.

Note that in our electronic structure calculations, we rely on experimentally
measured internal parameters $x$, $z_{\rm{Sr}}$, $z_{\rm{O}}$ and lattice
constants because they are well established for this
material and are probably still more accurate than can be obtained from DFT.

\paragraph{Calculation details}

The details of the computational method are described in our previous
papers \cite{AJY+06,AHY+07b,AYJ10,AKB22a} and here we only mention
several aspects. The band structure calculations were performed using
the fully relativistic linear muffin-tin orbital (LMTO) method
\cite{And75,book:AHY04}. This implementation of the LMTO method uses
four-component basis functions constructed by solving the Dirac
equation inside an atomic sphere \cite{NKA+83}. The
exchange-correlation functional of the generalized gradient
approximation (GGA)-type was used in the version of Perdew, Burke and
Ernzerhof \cite{PBE96}. The Brillouin zone integration was performed
using the improved tetrahedron method \cite{BJA94}. The basis
consisted of Ir and Sr $s$, $p$, $d$, and $f$; and O $s$, $p$, and $d$
LMTO's.

To take into account the electron-electron correlation effects we used
in this work the "relativistic" generalization of the rotationally
invariant version of the LSDA+$U$ method \cite{YAF03} which takes into
account that in the presence of spin-orbit coupling the occupation
matrix of localized electrons becomes non-diagonal in spin
indexes. Hubbard $U$ was considered as an external parameter and
varied from 0.65 eV to 3.65 eV.  We used in our calculations the value
of exchange Hund coupling $J_H$=0.65 eV obtained from constrained LSDA
calculations \cite{DBZ+84,PEE98}. Thus, the parameter $U_{eff}=U-J_H$,
which roughly determines the splitting between the lower and upper
Hubbard bands, varied between 0 eV and 3.0 eV. We adjusted the value
of $U$ to achieve the best agreement with the experiment.

In the RIXS process an electron is promoted from a core level to an
intermediate state, leaving a core hole. As a result, the electronic
structure of this state differs from that of the ground state. In
order to reproduce the experimental spectrum the self-consistent
calculations should be carried out including a core hole.  Usually, the
core-hole effect has no impact on the shape of XAS at the
$L_{2,3}$ edges of 5$d$ systems and just a minor effect on the XMCD
spectra at these edges \cite{book:AHY04}. However, the core hole has a
strong effect on the RIXS spectra in transition metal compounds
\cite{AKB22a,AKB22b}, therefore we take it into account.

\section{Electronic and magnetic structures}
\label{sec:bands}

\begin{figure}[tbp!]
\begin{center}
\includegraphics[width=0.8\columnwidth]{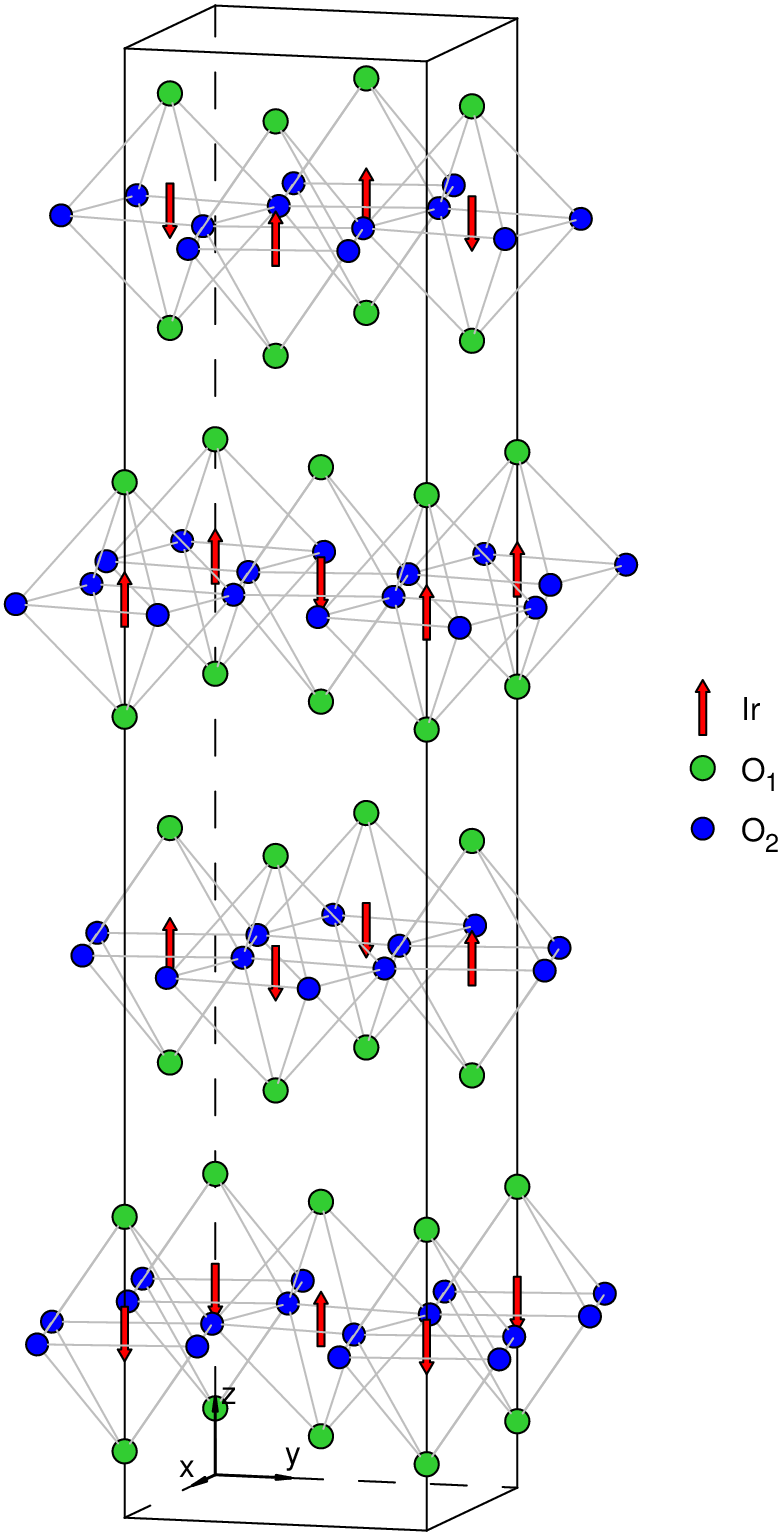}
\end{center}
\caption{\label{magn_str_SIO}(Color online) The AFM ordering with Ir moments
  parallel to $c$ axis used in GGA+SO+$U$ calculations. }
\end{figure}

\begin{figure}[tbp!]
\begin{center}
\includegraphics[width=0.99\columnwidth]{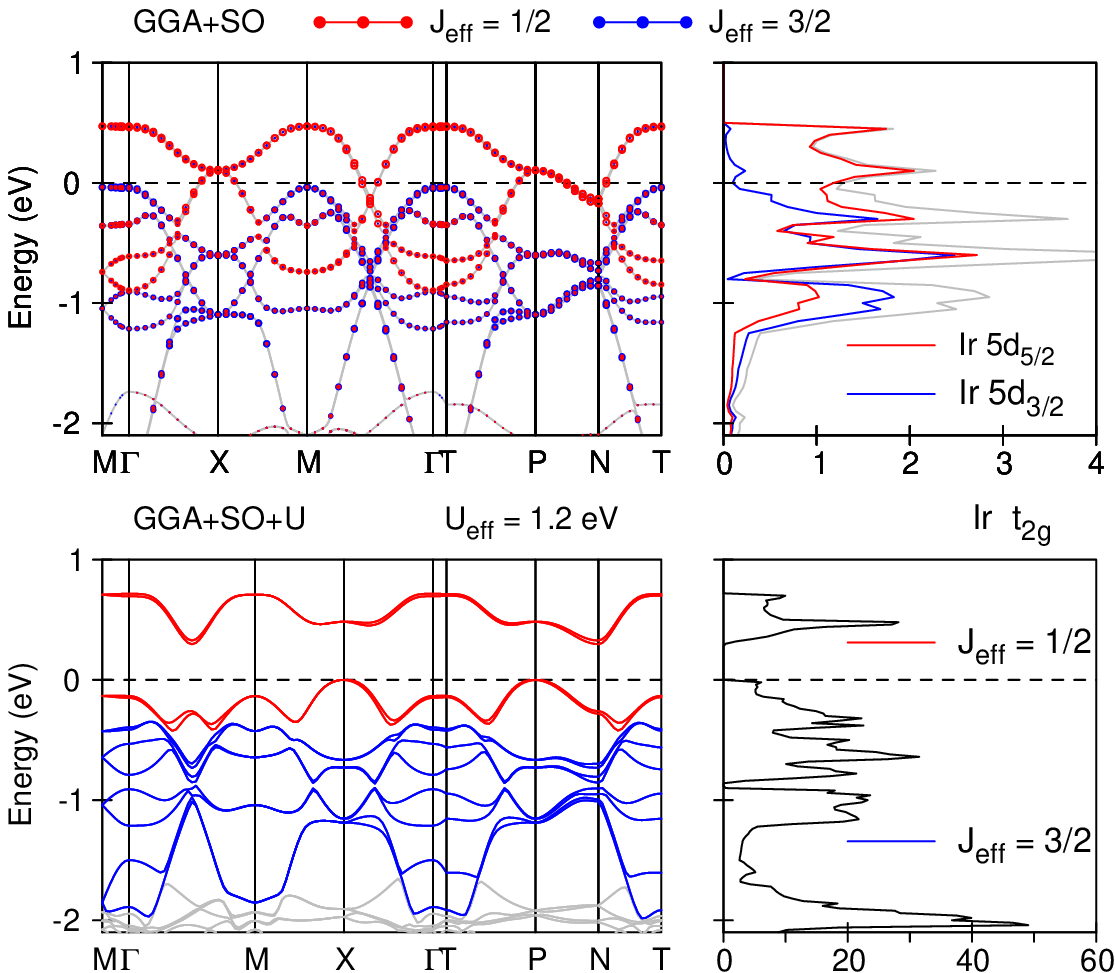}
\end{center}
\caption{\label{BND_Jeff_SIO}(Color online) The {\tg} energy band structure
  of Sr$_2$IrO$_4$ calculated in the fully relativistic Dirac GGA+SO
  approximation (the upper panel). The bands crossing the Fermi level which have
  almost pure $d_{5/2}$ character (open red circles) are formed by {\tg}
  states with $j_{eff}$ = 1/2. The lower panel presents the {\tg} energy bands
  calculated in the GGA+SO+$U$ approximation with $U_{eff}$ = 1.2 eV. }
\end{figure}

We performed GGA, GGA+SO, and GGA+SO+$U$ calculations of the
electronic and magnetic structures of Sr$_2$IrO$_4$ for the experimental
crystal structure \cite{SIN+95} and AFM ordering along the $c$ direction
(Fig. \ref{magn_str_SIO}). Although Sr$_2$IrO$_4$ has a canted AFM structure
\cite{KOK+09}, test calculations for a non-collinear AFM structure gave almost
identical optical, XA, XMCD, and RIXS spectra. Fig. \ref{BND_Jeff_SIO} (the
upper panel) presents the energy band structure of Sr$_2$IrO$_4$ in the energy
range of the Ir {\tg} manifold from $-$2.1 to 1.0 eV, calculated in the fully
relativistic Dirac GGA+SO approximation. The GGA+SO bands are presented by
circles proportional in size to their orbital character projected onto the
basis set of Ir $d_{3/2}$ (blue) and $d_{5/2}$ (red) states. The strong SOC
splits the {\tg} manifold into a lower $J_{eff}$ = 3/2 quartet and an upper
$J_{eff}$ = 1/2 doublet. The functions of the $J_{eff}$ = 3/2 quartet are
dominated by $d_{3/2}$ states with some weight of $d_{5/2}$ ones, which is
determined by the relative strength of SOC and crystal-field splitting. The
$J_{eff}$ = 1/2 functions are almost completely given by the linear combinations
of $d_{5/2}$ states. This allows one to identify the bands with pure $d_{5/2}$
character as originating from the $J_{eff}$ = 1/2 states. The GGA+SO approach
produces a metallic state in Sr$_2$IrO$_4$. The GGA+SO+$U$ approximation
shifts the occupied and empty {\tg} bands downward and upward, respectively,
by $U_{eff}$/2 producing a dielectric ground state (the lower panel 
of Fig. \ref{BND_Jeff_SIO}). The energy gap is increased with increasing Hubbard $U$.

\begin{figure}[tbp!]
\begin{center}
\includegraphics[width=0.99\columnwidth]{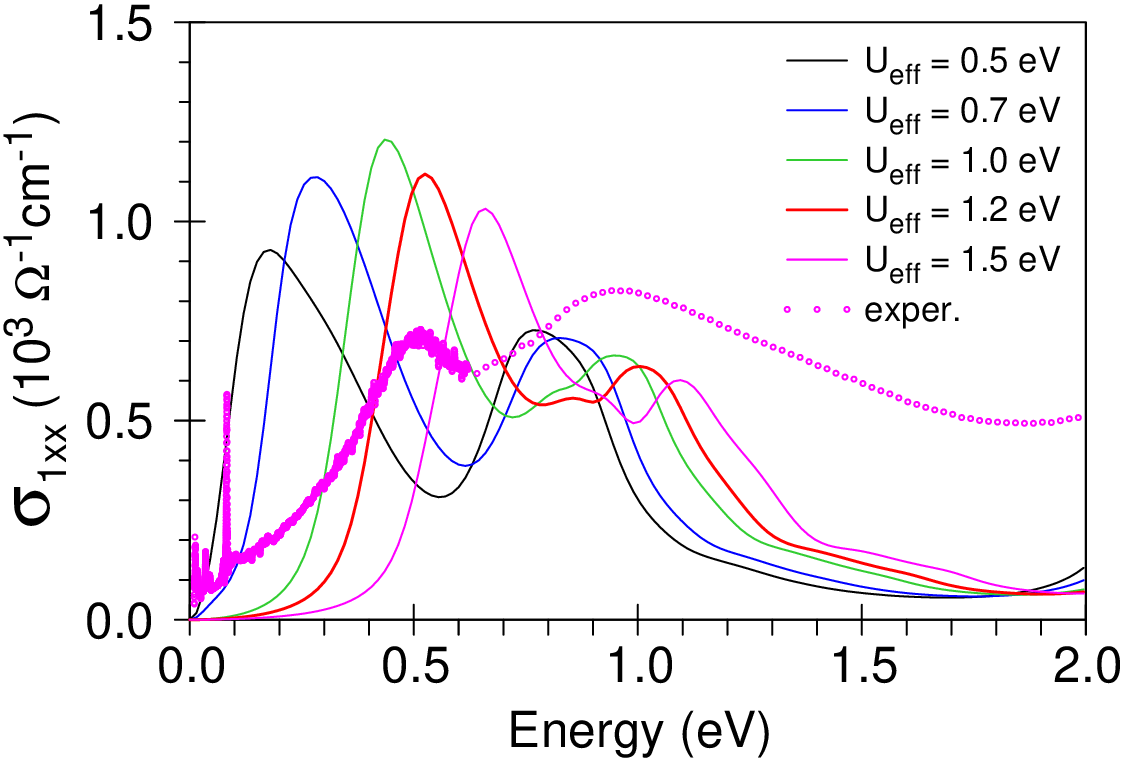}
\end{center}
\caption{\label{optic_SIO}(Color online) The experimentally measured real part
  of the optical conductivity (open magenta circles), $\sigma_{1xx}$,
  \cite{PYH+16a} (in 10$^3$ $\Omega^{-1}$ cm$^{-1}$) in Sr$_2$IrO$_4$ of the
  in-plane response compared with the theoretical spectra calculated in the
  GGA+SO+$U$ approximation for different $U_{eff}$ values. }
\end{figure}

Figure \ref{optic_SIO} presents the experimentally measured real part of the
optical conductivity (open magenta circles), $\sigma_{1xx}$, \cite{PYH+16a}
for the energy below 2 eV in Sr$_2$IrO$_4$
compared with the theoretical spectra calculated in the GGA+SO+$U$
approximation for different $U_{eff}$ values. The experimental optical
absorption consists of two peaks at around 0.5 and 1.0 eV. We found that the
low energy peak is derived from transitions between initial and final bands
formed by pure $J_{eff}$ = 1/2 states near band boundaries, e.g., around the X
point or the P-N high-symmetry line. The antiferromagnetic ordering of Ir
moments within the $ab$ plane stabilized by the on-site Coulomb repulsion $U$
causes a gap opening near the zone boundary between two pairs of bands
which show nearly parallel dispersion what insures high joint density of
states for the interband transitions responsible for the low energy peak. This is
in line with previous theoretical calculations \cite{ZHV13,PYH+16} and
experimental photoemission results \cite{WCW+13}. The high energy peak located
around 1 eV is dominated by a contribution from transitions with $J_{eff}$ = 3/2
initial states. Our calculations give the lower absorption peak about twice as
strong as the higher energy one, while in the experimental spectra the
strength is approximately the same for both. A similar trend was observed also
by Pr\"opper {\it et al.} \cite{PYH+16} and Kim {\it et al.} \cite{KKM12}. The
later authors relate this to an interband mixing of $J_{eff}$ = 3/2 and $J_{eff}$
= 1/2 states, which reflects the itinerancy of the system, i.e., the
hybridization of Ir 5$d$ states via neighboring oxygen 2$p$ states. The best
agreement between the calculated and experimentally measured energy positions
of the optical absorption peaks can be achieved for $U_{eff}$ = 1.2 eV.

\begin{figure}[tbp!]
\begin{center}
\includegraphics[width=0.89\columnwidth]{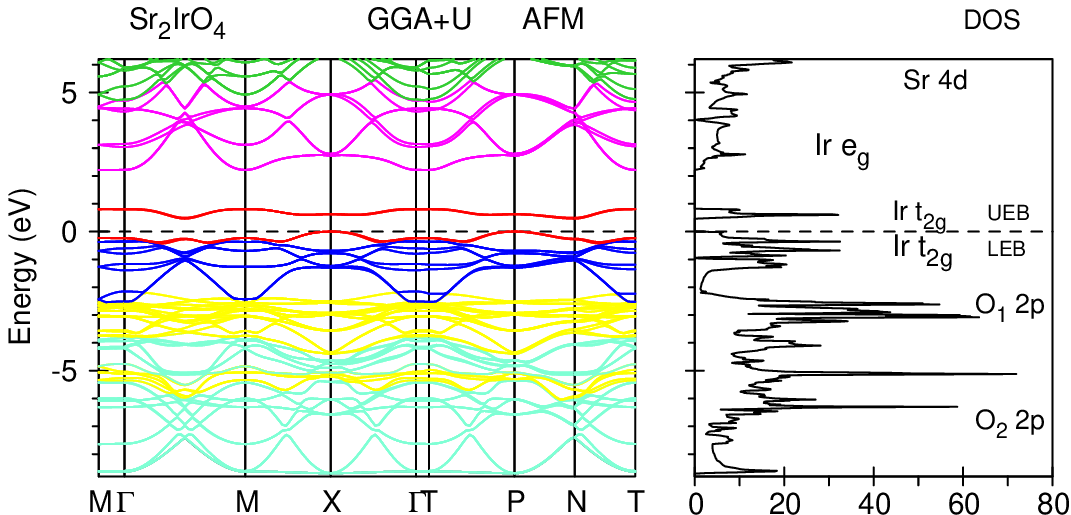}
\end{center}
\caption{\label{BND_SIO}(Color online) The energy band structure and
  total DOS [in states/(cell eV)] of Sr$_2$IrO$_4$ calculated with taking into
  account Coulomb correlations in the GGA+SO+$U$ approximation ($U_{eff}$ =
  1.2 eV). }
\end{figure}

\begin{figure}[tbp!]
\begin{center}
\includegraphics[width=0.9\columnwidth]{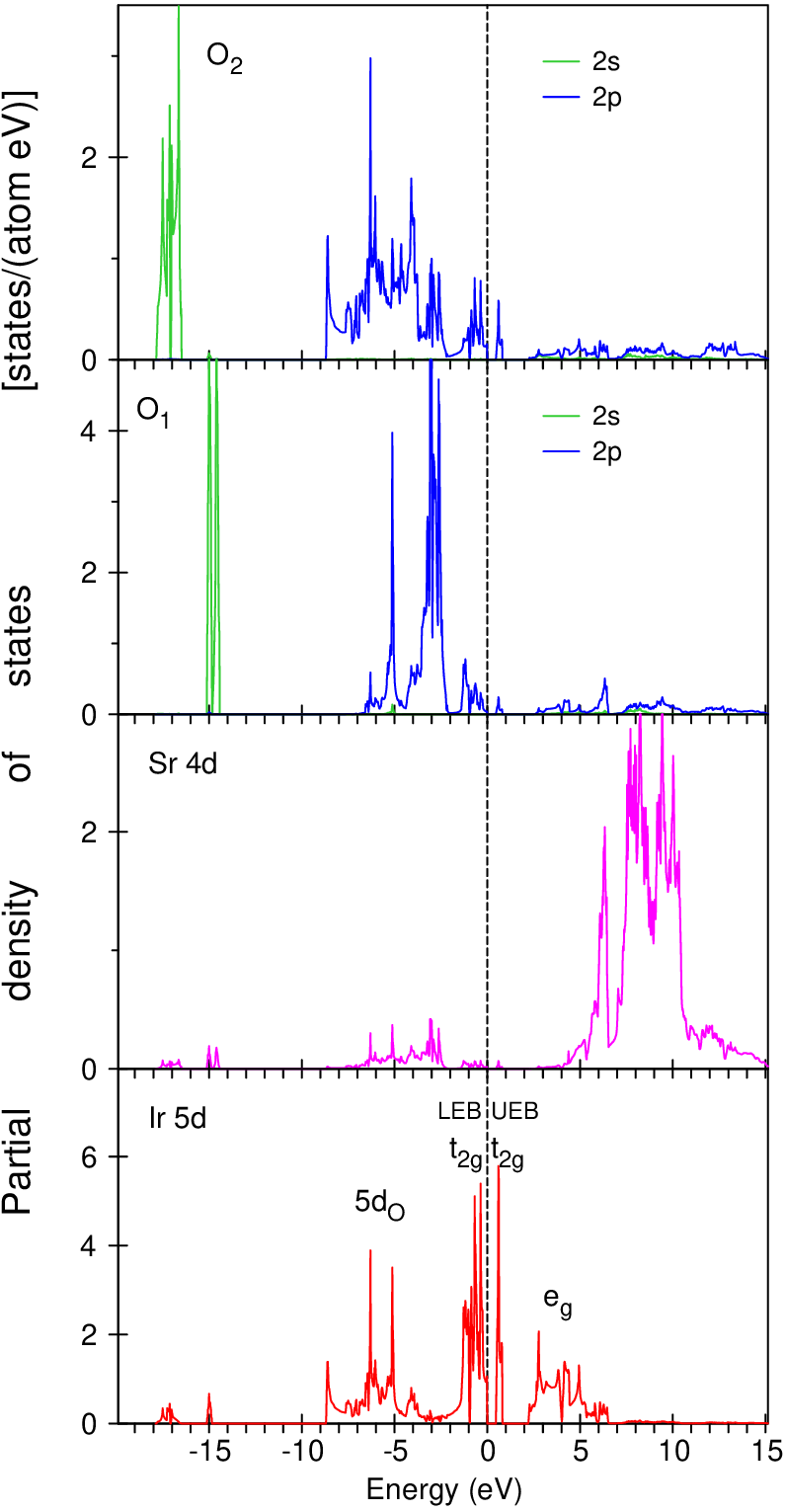}
\end{center}
\caption{\label{PDOS_SIO}(Color online) The partial DOSs for Sr$_2$IrO$_4$
  calculated in the GGA+SO+$U$ ($U_{eff}$= 1.2 eV) approximation. }
\end{figure}

Figures \ref{BND_SIO} and \ref{PDOS_SIO} present the energy band structure
and partial density of states (DOS), respectively, in Sr$_2$IrO$_4$ calculated
in the GGA+SO+$U$ approximation with $U_{eff}$ = 1.2 eV. Five electrons occupy
the {\tg}-type low energy band (LEB) manifold in the energy interval from
$-$1.5 eV to $E_F$ in Sr$_2$IrO$_4$. The empty {\tg} states [the upper energy
  band (UEB)] consist of one peak and occupy the energy range from 0.41 eV to
0.82 eV (see Fig. \ref{PDOS_SIO}). The {\eg}-type states of Ir are distributed
far above the Fermi level from 2.2 eV to 5.1 eV. The 4$d$ states of Sr ions
are mostly situated above the Fermi level from 5.3 to 10.8 eV. 

The electronic structure of apical O$_1$ and in-plane O$_2$ ions significantly
differ from each other. The apical O$_1$ 2$s$ states consist of two very
narrow peaks situated at $-$15.1 and $-$14.4 eV. The in-plane O$_2$ 2$s$ states
possess a relatively wider two peak structure from $-$17.9 to $-$16.5 eV. The
O$_1$ 2$p$ states are situated just below Ir LEB between $-$3.5 and $-$2.1 eV.
There is also a narrow peak at $-$5.2 eV. The in-plane O$_2$ 2$p$ states
occupy a relatively wide energy interval from $-$8.7 to $-$3.6 eV. The small
peaks in the close vicinity of the Fermi level from $-$1.5 eV to $E_F$ and from
0.41 eV to 0.82 eV are due to the strong hybridization between O 2$p$ and Ir
{\tg} LEB and UEB, respectively.

The occupation number of 5$d$ electrons in the Ir atomic sphere in Sr$_2$IrO$_4$
is equal to 6.3, which is much larger than the expected value of five
{\tg} electrons. The excessive charge is provided by the "tails" of oxygen
2$p$ states. These 5$d_{\rm{O}}$ states are located at the bottom of oxygen 2$p$
states from $-$8.7 eV to $-$3.1 eV and play an essential role in the RIXS
spectrum at the Ir $L_3$ edge (see Section IV). 

The theoretically calculated spin $M_s$, orbital $M_s$, and total $M_{total}$
magnetic moments using the GGA+SO+$U$ approach ($U_{eff}$ = 1.2 eV) for the
AFM solution are equal to 0.2647 {\mb}, 0.4447 {\mb}, and 0.7094 {\mb},
respectively. The spin and orbital magnetic moments at the Sr site are
relatively small ($M_s$ = 0.0007 {\mb} and $M_l$ = 0.0015 {\mb}). The
magnetic moments for apical O$_1$ ions are equal to $M_s$ = 0.0294 {\mb},
$M_l$ = 0.0254 {\mb}. For in-plane O$_2$ ions the magnetic moments almost vanish.

\section{I\lowercase{r} XMCD and RIXS spectra}
\label{sec:rixs}

\begin{figure}[tbp!]
\begin{center}
\includegraphics[width=0.9\columnwidth]{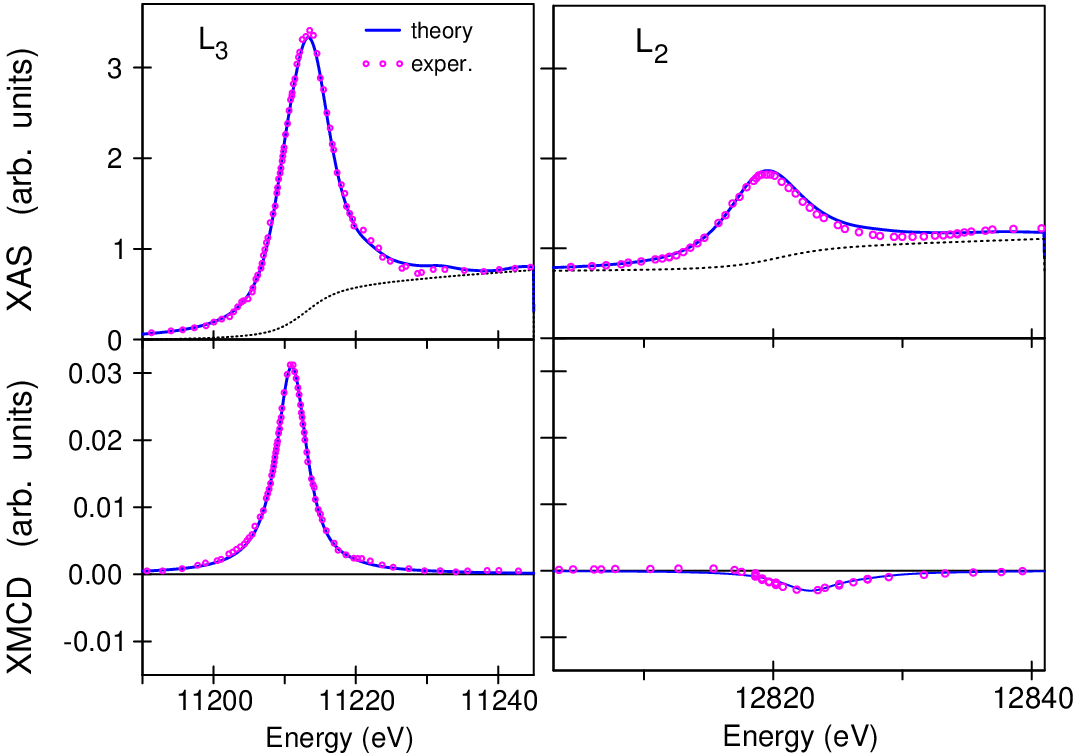}
\end{center}
\caption{\label{xmcd_Ir_SIO}(Color online) The experimental x-ray
  absorption (upper panels) and XMCD spectra (lower panels) at
  the Ir $L_{2,3}$ edges in the Sr$_2$IrO$_4$ thin film (magenta
  circles) \cite{HFZ+12} measured at 6 K under a 0.8 T magnetic field
  compared with the theoretically calculated spectra in the GGA+SO+$U$
  approximation (full blue curves). The dotted black curves in the upper
  panels show the background scattering intensity. }
\end{figure}

Figure \ref{xmcd_Ir_SIO} presents the experimentally measured x-ray absorption
spectra (the upper panel) and XMCD spectra (the lower panel) at the Ir $L_{2,3}$ edges
for Sr$_2$IrO$_4$ \cite{HFZ+12} (open circles) compared with the theoretically
calculated ones in the GGA+SO+$U$ ($U_{eff}$ = 1.2 eV) approximation (full
blue curves). The theoretically calculated Ir $L_{2,3}$ XA and XMCD spectra
are in good agreement with the experiment. The isotropic XA spectra are
dominated by the empty $e_g$ states with a smaller contribution from the empty
$t_{2g}$ orbitals at lower energy. The XMCD spectra, however, mainly come from the
$t_{2g}$ orbitals ($J_{eff}$ = 1/2). This results in a shift between the
maxima of the XA and XMCD spectra.

Due to the importance of SOC effects in iridates, it is natural to
quantify the strength of the SO interactions in these compounds. One method of
accomplishing this is provided by the x-ray absorption spectroscopy.  Van der
Laan and Thole showed that the so-called branching ratio BR = $I_{L_3}/I_{L_2}$
($I_{L_{2,3}}$ is the integrated intensity of the isotropic XAS at the
$L_{2,3}$ edges) is an important quantity in the study of 5$d$ oxides related to
the SO interaction \cite{LaTh88}. The BR is directly related to the ground-state
expectation value of the angular part of the spin-orbit coupling $<{\bf L
  \cdot S}>$ through BR = $(2 + r)/(1 - r)$, with $r$= $<{\bf L \cdot S}>/n_h$
and $n_h$ is the number of holes in 5$d$ states \cite{LaTh88}. As a result,
XAS provides a direct probe of SO interactions, which is complementary to
other techniques such as the magnetic susceptibility, electron paramagnetic
resonance, and Mossbauer spectroscopy (which probe SOC through the value of
the Lande g-factor). In the limit of negligible SOC effects the statistical
branching ratio BR = 2, and the $L_3$ white line is twice the size of the $L_2$
feature \cite{LaTh88}. The measured BR in Sr$_2$IrO$_4$ is close to 4.1
\cite{HFZ+12}, which differs significantly from the statistical BR = 2 in the
absence of orbital magnetization in 5$d$ states. A strong deviation from 2
indicates a strong coupling between the local orbital and spin
moments. Our DFT calculations produce BR = 3.56 for the GGA+SO+$U$ ($U_{eff}$
= 1.2 eV) approximation which is rather close to the experimental data of
Haskel {\it et al.}  \cite{HFZ+12}. It should be mentioned that the effect of
Coulomb correlations changes the energy band structure of transition metal
compounds in two ways. First, $d$ occupied states are shifted downward by
$U_{eff}$/2 and empty $d$ states are shifted upward by this value relative to
the Fermi energy. Second, the Coulomb correlations enhance an effective
spin-orbit coupling constant \cite{LAJ+08}. The relative influence of such
effect is increased in a row of 5$d \rightarrow 4d \rightarrow 3d$ transition
metal compounds due to an increase of Hubbard $U$ and a decrease of the atomic
SO coupling constant $\lambda_{SO}$.

\begin{figure}[tbp!]
\begin{center}
\includegraphics[width=0.9\columnwidth]{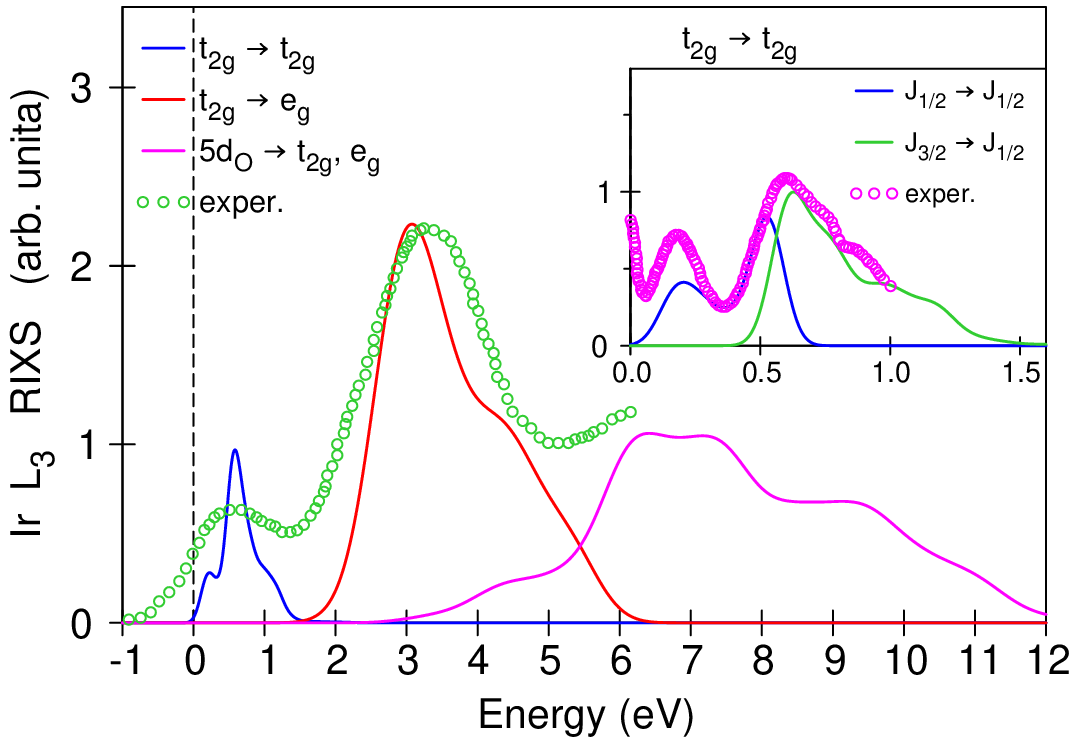}
\end{center}
\caption{\label{rixs_SIO}(Color online) The experimental RIXS spectrum (open
  green circles) measured by Ishii {\it et al.}  \cite{IJY+11} at the Ir $L_3$
  edge in Sr$_2$IrO$_4$ compared with the theoretically calculated one in the
  GGA+SO+$U$ approximation ($U_{eff}$=1.2 eV). Inset: The experimental RIXS spectrum
  (open magenta circles) measured by Clancy {\it et al.}  \cite{CGL+23} at the
  Ir $L_3$ edge for {\tg} $\rightarrow$ {\tg} transitions in Sr$_2$IrO$_4$
  compared with the theoretically calculated one in the GGA+SO+$U$ approach
  ($U_{eff}$=1.2 eV). }
\end{figure}

The RIXS spectra at Ir $L_{2,3}$ occur from a local excitation between the
filled and empty 5$d$ states. More precisely, the incoming photon excites a
2$p_{1/2}$ core electron ($L_2$ spectrum) or 2$p_{3/2}$ one ($L_3$ spectrum)
into an empty 5$d$ state what is followed by the de-excitation from the
occupied 5$d$ state into the core level. Because of the dipole selection
rules, apart from 6$s_{1/2}$-states (which have a small contribution to
RIXS due to relatively small 2$p$ $\rightarrow$ 6$s$ matrix elements
\cite{book:AHY04}) only 5$d_{3/2}$-states occur for $L_2$ RIXS, whereas for
$L_3$ RIXS 5$d_{5/2}$-states also contribute. Although the 2$p_{3/2}$
$\rightarrow$ 5$d_{3/2}$ radial matrix elements are only slightly smaller than
the 2$p_{3/2}$ $\rightarrow$ 5$d_{5/2}$ ones, the angular matrix
elements strongly suppress the 2$p_{3/2}$ $\rightarrow$ 5$d_{3/2}$
contribution \cite{book:AHY04}. Therefore, the RIXS spectrum at the Ir $L_3$ edge
can be viewed as interband transitions between 5$d_{5/2}$ states.

Figure \ref{rixs_SIO} shows the experimental RIXS spectrum (open green circles)
measured by Ishii {\it et al.} \cite{IJY+11} at the Ir $L_3$ edge in
Sr$_2$IrO$_4$ compared with the theoretically calculated one in the GGA+SO+$U$
approximation ($U_{eff}$=1.2 eV). The Ir $L_3$ RIXS spectrum consists of two peaks
below 5 eV. We found that the low energy peak corresponds to intra-{\tg}
excitations. This fine structure has a two peak structure in our calculations
but the measurements of Ishii {\it et al.} \cite{IJY+11} show only one peak. 
However, Clancy {\it et al.} \cite{CGL+23} using higher resolution
were able to distinguish two peaks (see the inset in
Fig. \ref{rixs_SIO}). The low energy peak at 0.5 eV is due to interband
transitions between occupied and empty Ir $J_{eff}$ = 1/2 states (the red curves in
the lower panel of Fig. \ref{BND_Jeff_SIO}). These transitions also
contribute to the second high energy peak at around 0.7 eV together with
$J_{3/2}$ $\rightarrow$ $J_{1/2}$ transitions (the green curve in the inset of
Fig. \ref{rixs_SIO}). The intensive peak at around 3.4 eV (the red curve in Fig.
\ref{rixs_SIO}) is due to $\tg \rightarrow \eg$ transitions. The next fine
structure from 4.5 eV to 12 eV (the magenta curve) is due to 5$d_{\rm{O}}$
$\rightarrow$ {\tg}, {\eg} transitions.

\begin{figure}[tbp!]
\begin{center}
\includegraphics[width=0.9\columnwidth]{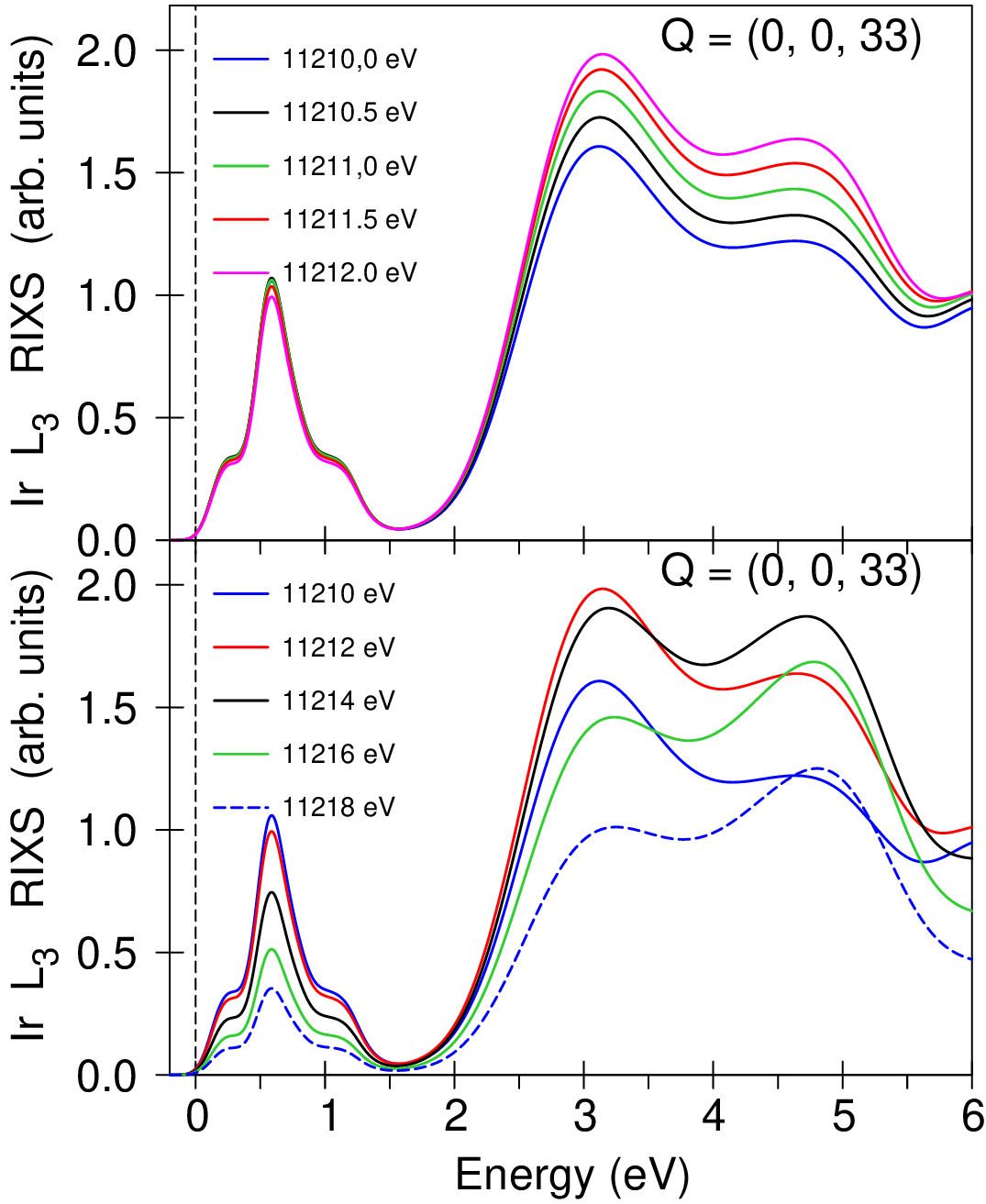}
\end{center}
\caption{\label{rixs_Ir_Ei}(Color online) The RIXS spectra as a function of
  incident photon energy $E_i$ calculated at the Ir $L_3$ edge in
  Sr$_2$IrO$_4$ with the momentum transfer vector {\bf Q} = (0, 0, 33). }
\end{figure}

Figure \ref{rixs_Ir_Ei} shows the Ir $L_3$ RIXS spectrum as a function of incident
photon energy $E_i$ above the corresponding edge with the momentum transfer vector
{\bf Q} = (0, 0, 33). We found that the low energy fine structure
corresponding to intra-{\tg} excitations is slightly decreased when the
incident photon energy changes in the interval of 2 eV above the Ir $L_3$ edge,
while the high energy peak corresponding to the $\tg \rightarrow \eg$ transitions
is monotonically increased (the upper panel of Fig. \ref{rixs_Ir_Ei}). A similar trend 
was observed in the measurements of Ishii {\it et al.}  \cite{IJY+11}. They measured 
the RIXS spectra of Sr$_2$IrO$_4$ at two representative incident photon energies in the
interval of 3 eV and discovered that with increasing $E_i$ the low energy peak
is decreased but the high energy peak is increased [for the same momentum
transfer vector {\bf Q} = (0, 0, 33)]. The lower panel of Fig. \ref{rixs_Ir_Ei} 
shows the Ir $L_3$ RIXS spectrum as a function of incident photon energy in the larger 
energy interval of 18 eV. With increasing $E_i$ the low energy peak is 
steadily decreased, but the high energy peak shows more complex behavior. 
First, the intensity of the peak is increased, but then decreased with changing 
the relative intensity of the two peaks at 3 and 5 eV .

\begin{figure}[tbp!]
\begin{center}
\includegraphics[width=0.9\columnwidth]{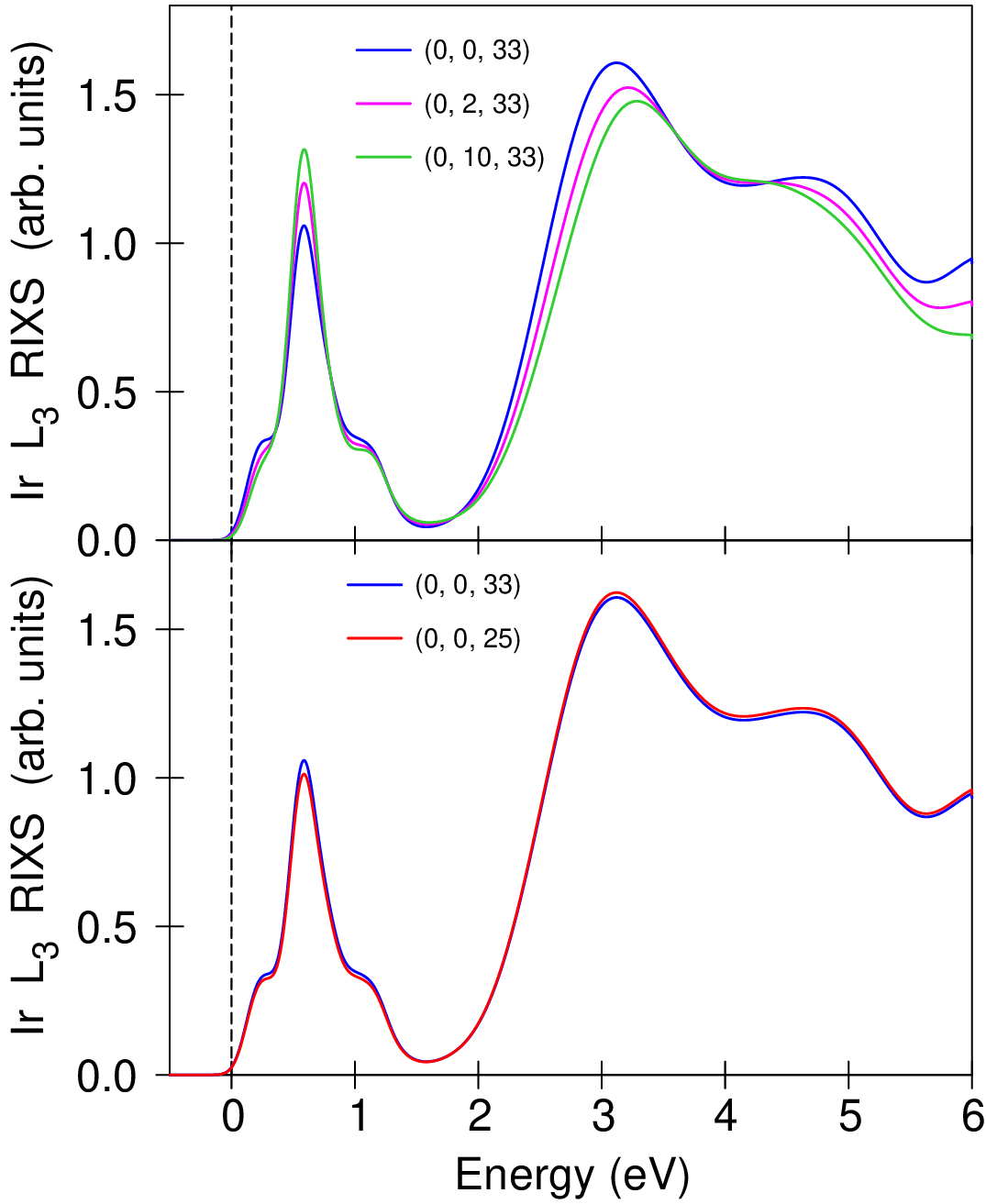}
\end{center}
\caption{\label{rixs_Ir_Qxy}(Color online) The RIXS spectra at the Ir $L_3$ edge
  in Sr$_2$IrO$_4$ calculated as a function of $Q_y$ (the upper panel) and
  $Q_z$ (the lower panel) with the momentum transfer vector {\bf Q} = (0, Q$_y$,
  Q$_z$) for incident photon energy $\hbar \omega_{in}$ = 11210 eV. }
\end{figure}

It is widely believed that $d-d$ excitations show only small momentum
transfer vector {\bf Q} dependence in 5$d$ transition metal compounds
\cite{LKH+12,KTD+20}. In particular, Sr$_2$IrO$_4$ has a layered-perovskite
structure, therefore, the momentum dependence along the $c$ axis is expected to be
small, as in high-$T_c$ cuprates \cite{ITE+05}. Indeed, as we see in the lower
panel of Fig. \ref{rixs_Ir_Qxy}, the RIXS spectra are almost identical for the
transfer vectors {\bf Q} = (0, 0, 33) and (0, 0, 25). Similar dependence was
observed also in the measurements of Ishii {\it et al.}  \cite{IJY+11}. The
upper panel of Fig. \ref{rixs_Ir_Qxy} shows the RIXS spectra at the Ir $L_3$
edge in Sr$_2$IrO$_4$ calculated as a function of $Q_y$ with the momentum
transfer vector {\bf Q} = (0, Q$_y$, 33) for incident photon energy $\hbar
\omega_{in}$ = 11210 eV. We found that with increasing Q$_y$ the first low
energy peak is increased and the high energy fine structure is
decreased. Analyzing Fig. \ref{rixs_Ir_Qxy} we can conclude that the momentum
dependence of the excitations in Sr$_2$IrO$_4$ is rather small as it was
earlier observed in other iridates such as Sr$_3$CuIrO$_6$ \cite{LKH+12} or
In$_2$Ir$_2$O$_7$ \cite{KTD+20}.

\section{O XAS and RIXS spectra}
\label{sec:rixs}

\begin{figure}[tbp!]
\begin{center}
\includegraphics[width=0.9\columnwidth]{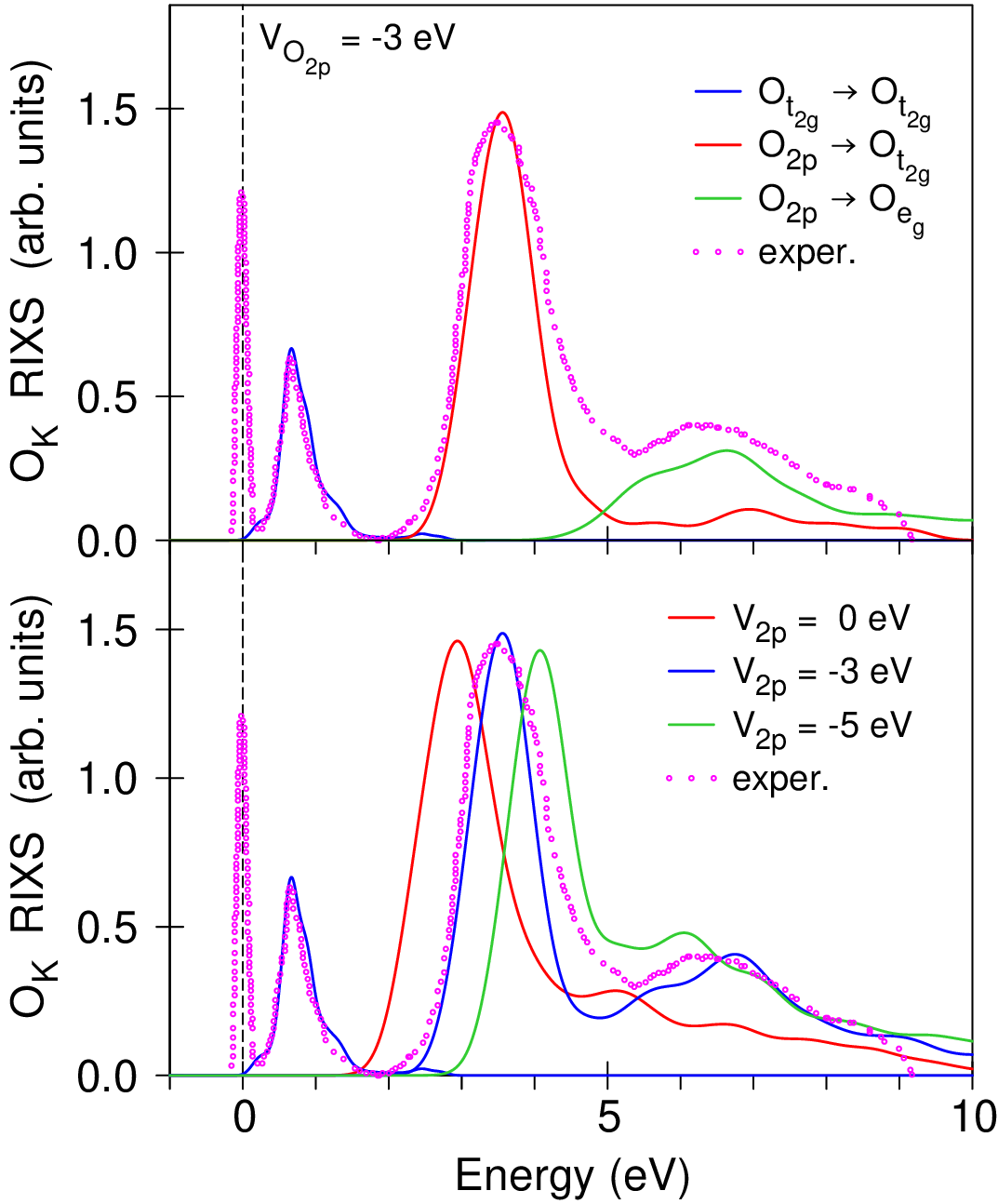}
\end{center}
\caption{\label{rixs_O_SIO}(Color online) (The upper panel) The experimental
  RIXS spectrum (open magenta circles) measured by Liu {\it et al.}
  \cite{LDL+15} at the O $K$ edge in Sr$_2$IrO$_4$ compared with the
  theoretically calculated ones in the GGA+SO+$U$+SIC approximation ($U_{eff}$
  = 1.2 eV, $V_{O_{2p}}$ = $-$3 eV). (The lower panel) The experimental RIXS
  spectrum (open magenta circles) measured by Liu {\it et al.} \cite{LDL+15}
  at the O $K$ edge in Sr$_2$IrO$_4$ compared with the theoretically
  calculated ones in the GGA+SO+$U$+SIC approach ($U_{eff}$ = 1.2 eV) for
  different parameters $V_l$. }
\end{figure}

The RIXS spectra at the O $K$ edge in Sr$_2$IrO$_4$ were measured by
Liu {\it et al.} \cite{LDL+15}, Lu {\it et al.} \cite{LOH+18}, Paris {\it et
  al.}  \cite{PTP+20}, and Kim {\it et al.} \cite{KDK+23}. The last three
investigations concentrate on the analysis of low energy excitations below 1.2
eV. Liu {\it et al.} \cite{LDL+15} present the RIXS spectrum up to 12 eV using
circular and $\pi$ polarizations of the incident beam. The O $K$ RIXS spectrum
consists of a peak centered at zero energy loss, which comprises the elastic
line and other low-energy features such as phonons, magnons, etc., and three
major inelastic excitations at 0.7 eV, 3.5 eV, and around 6.2. We found that
the first low energy feature is due to the interband transitions between occupied
and empty O$_{\tg}$ states, which appear as a result of the strong hybridization
between oxygen 2$p$ states with Ir {\tg} LEB and UEB in the close vicinity of the
Fermi level (see Fig. \ref{PDOS_SIO}), therefore, the oxygen $K$ RIXS spectroscopy
can be used for the estimation of the energy band gap and positions of Ir 5$d$
Hubbard bands. The next two peaks at around 3.5 and 6.2 eV reflect the
interband transitions from the occupied O 2$p$ states and the empty oxygen
states which originate from the hybridization with Ir {\tg} and {\eg}
states, respectively. We found that the theory reproduces well the shape and
energy position of the low energy feature, but the second and the third peaks are
shifted towards smaller energy in comparison with the experimental
measurements. It means that the DFT calculations cannot produce the correct
energy position of the oxygen 2$p$ bands. These bands are almost fully
occupied in Sr$_2$IrO$_4$, therefore, they cannot be described by
the GGA+$U$ method. To reproduce the correct energy position of
the oxygen 2$p$ band in Sr$_2$IrO$_4$ we used a self-interaction-like
correction procedure as proposed by Kaneko {\it et al.} \cite{KTK+13},
where the valence bands are shifted downwards by adding a SIC-like
orbital-dependent potential $V_l$ into the Hamiltonian. We used $V_l$ as a
parameter an adjusted it to produce the correct energy position of the oxygen 2$p$
bands. We found that the best agreement with the experiment can be achieved
for $V_{O_{2p}}$ = $-$3.0 eV (see the lower panel of
Fig.~\ref{rixs_O_SIO}).

\begin{figure}[tbp!]
\begin{center}
\includegraphics[width=0.9\columnwidth]{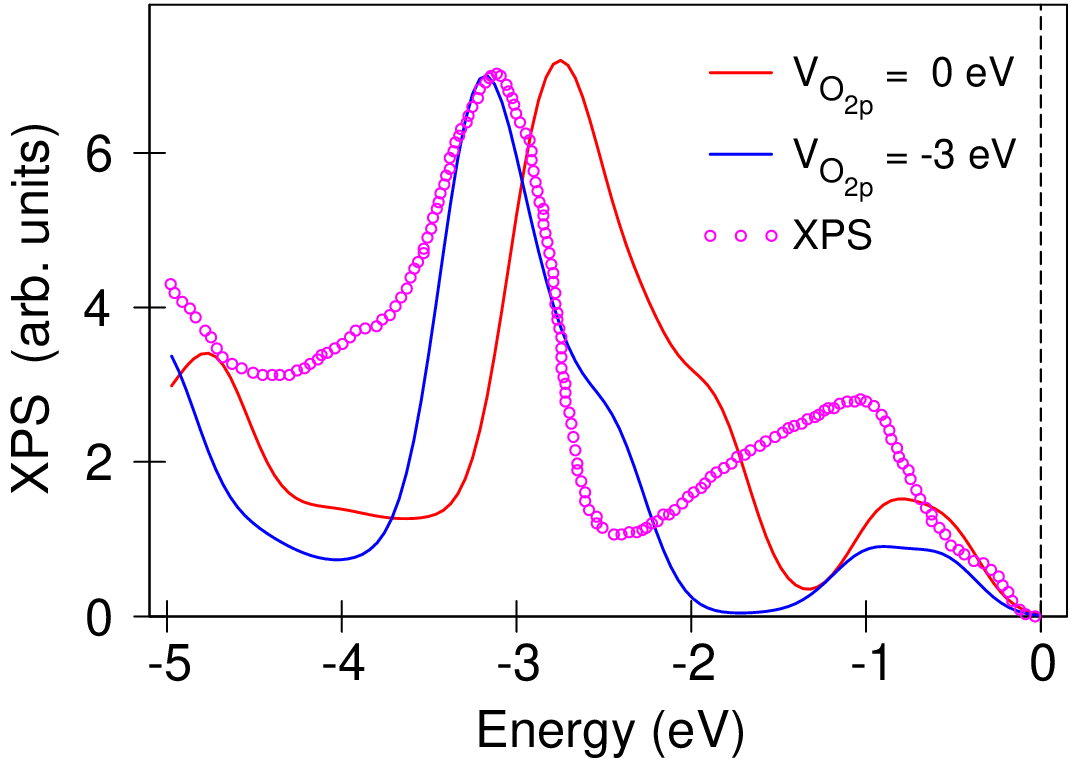}
\end{center}
\caption{\label{XPS_SIO}(Color online) The valence band photoemission spectrum of
  Sr$_2$IrO$_4$ \cite{KTT+13} compared with total DOS calculated in the
  GGA+SO+$U$+SIC approach ($U_{eff}$ = 1.2 eV) for different parameters
  $V_l$. }
\end{figure}

Figure \ref{XPS_SIO} presents the valence band photoemission spectrum of
Sr$_2$IrO$_4$ \cite{KTT+13} compared with total DOS calculated in the
GGA+SO+$U$+SIC approach ($U_{eff}$ = 1.2 eV). Two dominant peaks at around
$-$1 and $-$3.2 eV are observed, which might be attributed to the
photoemmision from Ir {\tg} and oxygen 2$p$ states, respectively
\cite{KTT+13}. Like in the case of O $K$ RIXS spectrum, the GGA+SO+$U$
approach cannot reproduce the correct energy position of the peak at $-$3.2
eV. However, the SIC-like approach with $V_{O_{2p}}$ = $-$3 eV improves the
situation.

\begin{figure}[tbp!]
\begin{center}
\includegraphics[width=0.9\columnwidth]{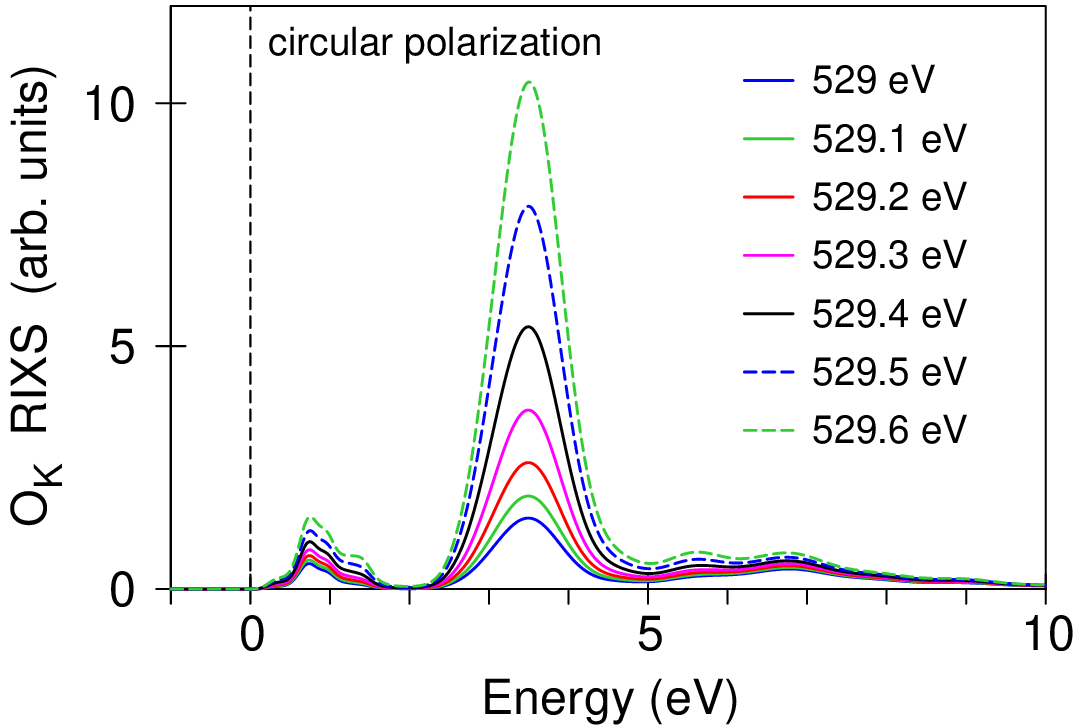}
\end{center}
\caption{\label{rixs_O_Ei}(Color online) The RIXS spectra as a function of
  incident photon energy calculated at the O $K$ edge in Sr$_2$IrO$_4$ with
  circular polarization. }
\end{figure}

\begin{figure}[tbp!]
\begin{center}
\includegraphics[width=0.9\columnwidth]{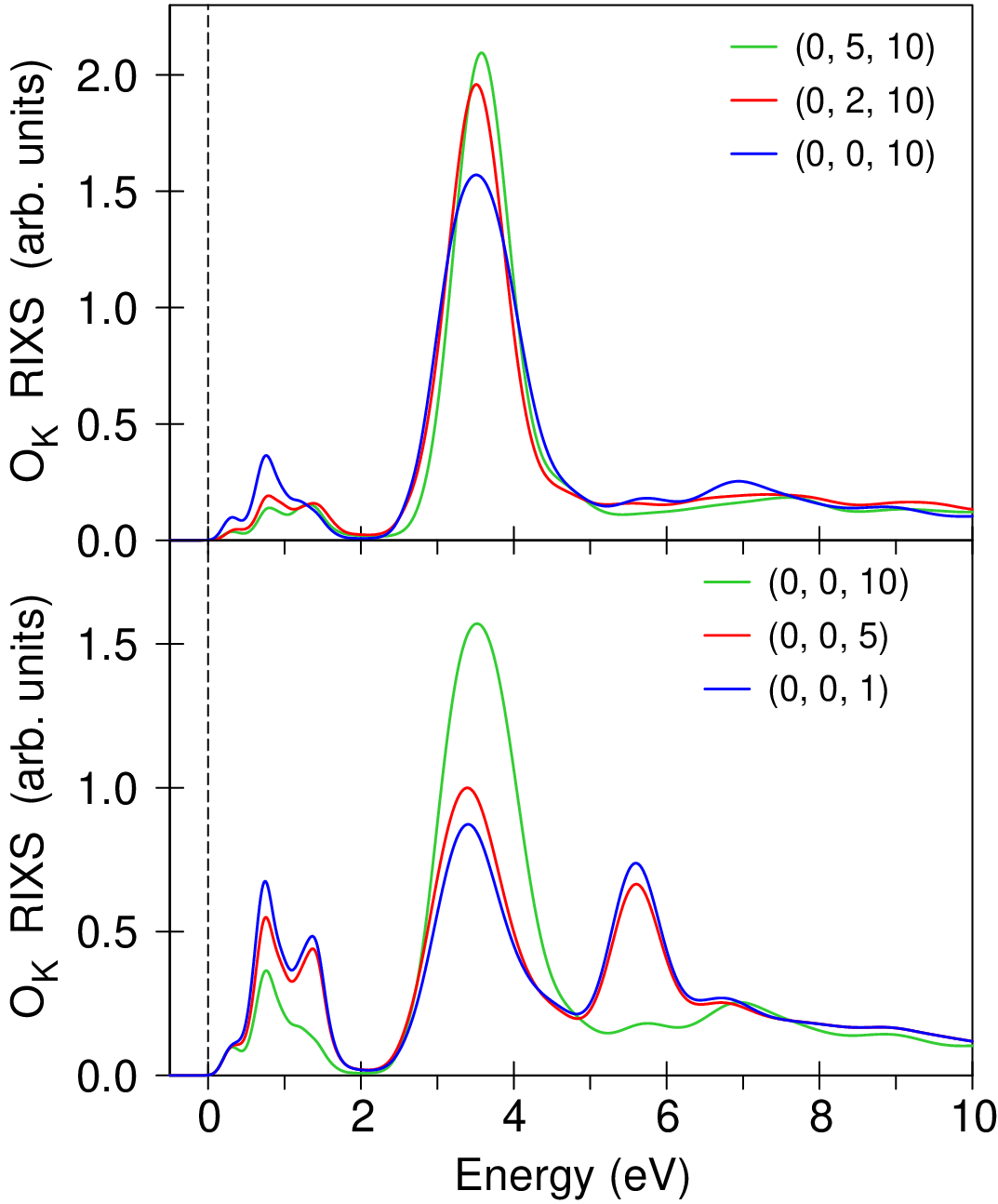}
\end{center}
\caption{\label{rixs_O_Qxy}(Color online) The RIXS spectra at the O $K$ edge in
  Sr$_2$IrO$_4$ calculated as a function of $Q_y$ (the upper panel) and $Q_z$
  (the lower panel) with the momentum transfer vector {\bf Q} = (0, Q$_y$,
  Q$_z$) for incident photon energy $\hbar \omega_{in}$ = 529 eV. }
\end{figure}

Figure \ref{rixs_O_Ei} presents the RIXS spectra as a function of incident photon
energy $E_i$ calculated at the O $K$ edge in Sr$_2$IrO$_4$ with circular
polarization. We found much stronger dependence on the incident photon
energy in the case of the O $K$ RIXS spectrum in comparison with the
corresponding dependence at the Ir $L_3$ edge (compare Figs. \ref {rixs_Ir_Ei}
and \ref{rixs_O_Ei}). With increasing the incident photon energy both peaks at
0.7 eV and 3.5 eV are increased, the later one is increased dramatically. And
this occurs in a small energy interval for $E_i$ of 0.6 eV.

Figure \ref{rixs_O_Qxy} shows the RIXS spectra at the O $K$ edge in Sr$_2$IrO$_4$
calculated as a function of $Q_y$ (the upper panel) and $Q_z$ (the lower
panel) with the momentum transfer vector {\bf Q} = (0, Q$_y$, Q$_z$). With
decreasing parameters Q$_x$ and Q$_z$ the intensity of the major peak at 3.5
eV is decreased but the low energy peak at 0.7 eV is increased. There is also
a strong change in the shape of the low energy peak at 0.7 eV and the third peak
at around 6.2 eV with the change of the parameter Q$_z$.

\begin{figure}[tbp!]
\begin{center}
\includegraphics[width=0.9\columnwidth]{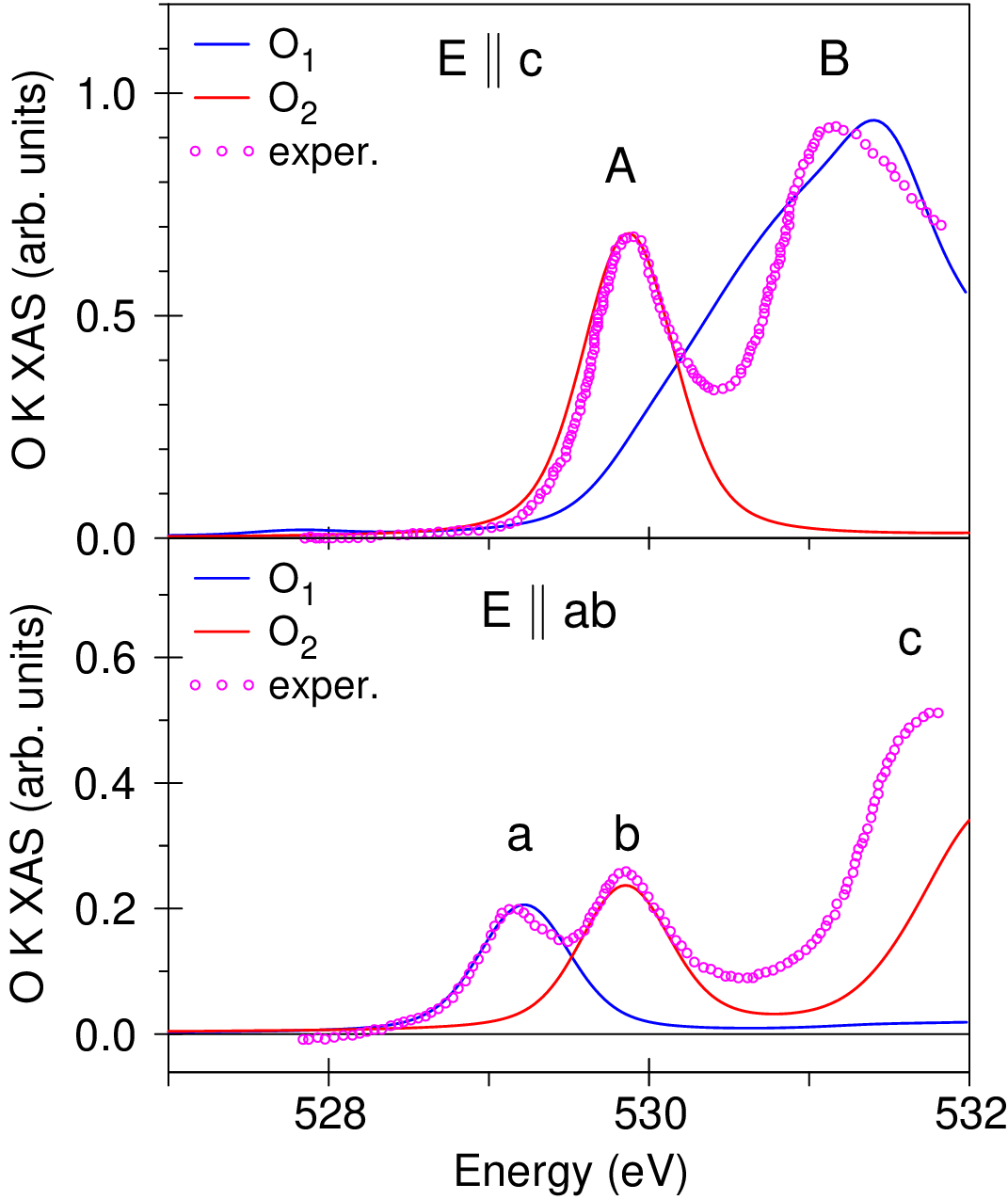}
\end{center}
\caption{\label{XAS_O_SIO}(Color online) The experimental O $K$ polarization
  dependent XA spectra (open magenta circles) \cite{KJM+08} in Sr$_2$IrO$_4$
  compared with the theoretically calculated ones in the GGA+SO+$U$+SIC
  approach ($U_{eff}$ = 1.2 eV, $V_{O_{2p}}$ = $-$3 eV). }
\end{figure}

Figure \ref{XAS_O_SIO} presents the experimental O $K$ polarization dependent
x-ray absorption spectra (open magenta circles) \cite{KJM+08} compared with
the theoretically calculated ones in the GGA+SO+$U$+SIC approach ($U_{eff}$ =
1.2 eV, $V_{O_{2p}}$ = $-$3 eV). Due to the quasi-two-dimensional structure of
Sr$_2$IrO$_4$ there is strong anisotropy in the x-ray absorption
spectra. There are two small peaks $a$ and $b$ at 529.1 and 529.9 eV and
larger peak $c$ at around 531.8 eV for the {\bf E} $\perp$ $c$ polarization and
only two peaks $A$ and $B$ for {\bf E} $\parallel$ $c$ at 529.9 and 531.3
eV. We found that the low energy peak $a$ for {\bf E} $\perp$ $c$ and the large peak
$B$ for {\bf E} $\parallel$ $c$ are derived from the apical oxygens O$_1$. The peaks
$b$ and $A$ are due to the 1$s$ $\rightarrow$ 2$p$ x-ray absorption on the
in-plane O$_2$ oxygens. The theory relatively well reproduces the
experimentally measured XA spectra.

\section{Conclusions}

The electronic and magnetic structures of Sr$_2$IrO$_4$ were investigated
theoretically in the frame of the fully relativistic spin-polarized Dirac LMTO
band-structure method in order to understand the importance of Coulomb
interaction and spin-orbit coupling. We also present comprehensive
theoretical calculations of the XA, XMCD, and RIXS spectra at the Ir $L_{2,3}$
and oxygen $K$ edges.

The strong SOC splits the {\tg} manifold into a lower $J_{eff}$ = 3/2 quartet
and an upper $J_{eff}$ = 1/2 doublet. The functions of the $J_{eff}$ = 3/2
quartet are dominated by $d_{3/2}$ states with some weight of $d_{5/2}$ ones,
which is determined by the relative strength of the SOC and crystal-field
splitting. The $J_{eff}$ = 1/2 functions are almost completely given by linear
combinations of $d_{5/2}$ states. This allows one to identify the bands with
pure $d_{5/2}$ character as originating from the $J_{eff}$ = 1/2 states. The
GGA+SO approach produces a metallic state in Sr$_2$IrO$_4$. The GGA+SO+$U$
approximation shifts the occupied and empty {\tg} bands downward and upward,
respectively, by $U_{eff}$/2 producing a dielectric ground state in
Sr$_2$IrO$_4$.

We found that the best agreement between the calculated and experimentally
measured optical conductivity spectrum can be achieved for $U_{eff}$ = 1.2 eV.
The experimental optical absorption consists of two peaks at around 0.5 and
1.0 eV. The low energy peak is derived from transitions between initial and
final bands formed by pure $J_{eff}$ = 1/2 states near band boundaries. The
high energy peak located around 1 eV is dominated by a contribution from
transitions with $J_{eff}$ = 3/2 initial states.

The theoretically calculated Ir $L_{2,3}$ XAS and XMCD spectra are in good
agreement with the experiment. The isotropic XA spectra are dominated by the
empty $e_g$ states with a smaller contribution from the empty $t_{2g}$
orbitals at lower energy. The XMCD spectra, however, mainly come from the $t_{2g}$
orbitals ($J_{eff}$ = 1/2). This results in a shift between the maxima of the
XA and XMCD spectra. The ratio BR = $I_{L_3}/I_{L_2}$ is an important
quantity in the study of the SO interaction in 5$d$ oxides. It is directly related
to the ground-state expectation value of the angular part of the spin-orbit
coupling. Our DFT calculations produce BR = 3.56 which is rather close to the
experimental value of 4.1 \cite{HFZ+12}. These values differ significantly from
the statistical BR = 2 in the absence of orbital magnetization in 5$d$
states. A strong deviation from 2 indicates a strong coupling
between the local orbital and spin moments.

The Ir $L_3$ RIXS spectrum consists of two peaks below 5 eV. We found that the
low energy peak corresponds to intra-{\tg} excitations. This fine structure
has a two peak structure. The low energy peak at 0.5 eV is due to interband
transitions between occupied and empty Ir $J_{eff}$ = 1/2 states. These
transitions also contribute to the second high energy peak at around 0.7
eV together with $J_{3/2}$ $\rightarrow$ $J_{1/2}$ transitions. The
intensive peak at around 3.4 eV is due to $\tg \rightarrow \eg$
transitions. The next fine structure from 4.5 eV to 12 eV is due to
5$d_{\rm{O}}$ $\rightarrow$ {\tg}, {\eg} transitions.

We investigated theoretically the influence of the incident photon energy
$E_i$ and the momentum transfer vector {\bf Q} on the shape of the Ir $L_3$
RIXS spectrum. We found that with inreasing of $E_i$ in the interval of 2 eV
above the Ir $L_3$ edge the low energy fine structure corresponding to
intra-{\tg} excitations is decreased and the high energy peak corresponding to
the $\tg \rightarrow \eg$ transitions is monotonically increased in agreement
with the measurements of Ishii {\it et al.} \cite{IJY+11}. The momentum
dependence of Ir $L_3$ RIXS was found to be relatively small. With
increasing Q$_y$ the first low energy peak is increased and the high energy
fine structure is slightly decreased. The variation of parameter Q$_z$ almost
does not influence the RIXS spectrum at the Ir $L_3$ edge.

The RIXS spectrum of Sr$_2$IrO$_4$ at the O $K$ edge consists of three
major inelastic excitations at 0.7 eV, 3.5 eV, and around 6.2 eV. We found
that the first low energy feature is due to the interband transitions between
occupied and empty O$_{\tg}$ states, which appear as a result of the strong
hybridization between oxygen 2$p$ states with Ir {\tg} LEB and UEB in the close
vicinity of the Fermi level. The next two peaks at around 3.5 and 6.2 eV
reflect the interband transitions from the occupied O 2$p$ states to the empty
oxygen states which originate from the hybridization with Ir {\tg} and
{\eg} states, respectively. We found that the theory reproduces well the shape
and energy position of the low energy feature, but the second and the third peaks
are shifted towards smaller energy in comparison with the experimental
measurements. It means that the DFT calculations cannot produce the correct
position of oxygen 2$p$ bands. To reproduce the correct energy position of the
oxygen 2$p$ band in Sr$_2$IrO$_4$ we used a self-interaction-like correction
procedure. We added a SIC-like orbital-dependent potential $V_l$ into the
Hamiltonian and found that the best agreement with the experiment can be
achieved for $V_{O_{2p}}$ = $-$3.0 eV. We found that the dependence of
the RIXS spectrum at the oxygen $K$ edge on the incident photon energy and the
momentum transfer vector {\bf Q} is much stronger in comparison with the
correspondent dependency at the Ir $L_3$ edge.

Due to the quasi-two-dimensional structure of Sr$_2$IrO$_4$ there is
strong anisotropy in the x-ray absorption spectra at the oxygen $K$
edge. There are two small peaks $a$ and $b$ at 529.1 and 529.9 eV and
a larger peak $c$ at around 531.8 eV for the {\bf E} $\perp$ $c$
polarization and only two peaks $A$ and $B$ for {\bf E} $\parallel$
$c$ at 529.9 and 531.3 eV. We found that the low energy peak $a$ for
{\bf E} $\perp$ $c$ and the large peak $B$ for {\bf E} $\parallel$ $c$
are derived from the apical oxygens O$_1$. The peaks $b$ and $A$ are
due to the 1$s$ $\rightarrow$ 2$p$ x-ray absorption on the in-plane
O$_2$ oxygens. The theory relatively well reproduces the
experimentally measured XA spectra.

\section*{Acknowledgments}

We are thankful to Dr. Alexander Yaresko from the Max-Planck-Institute FKF in
Stuttgart for helpful discussions.
 

\begin{thebibliography}{78}
\expandafter\ifx\csname natexlab\endcsname\relax\def\natexlab#1{#1}\fi
\expandafter\ifx\csname bibnamefont\endcsname\relax
  \def\bibnamefont#1{#1}\fi
\expandafter\ifx\csname bibfnamefont\endcsname\relax
  \def\bibfnamefont#1{#1}\fi
\expandafter\ifx\csname citenamefont\endcsname\relax
  \def\citenamefont#1{#1}\fi
\expandafter\ifx\csname url\endcsname\relax
  \def\url#1{\texttt{#1}}\fi
\expandafter\ifx\csname urlprefix\endcsname\relax\def\urlprefix{URL }\fi
\providecommand{\bibinfo}[2]{#2}
\providecommand{\eprint}[2][]{\url{#2}}

\bibitem[{\citenamefont{Jackeli and Khaliullin}(2009)}]{JaKh09}
\bibinfo{author}{\bibfnamefont{G.}~\bibnamefont{Jackeli}} \bibnamefont{and}
  \bibinfo{author}{\bibfnamefont{G.}~\bibnamefont{Khaliullin}},
  \bibinfo{journal}{Phys. Rev. Lett.} \textbf{\bibinfo{volume}{102}},
  \bibinfo{pages}{017205} (\bibinfo{year}{2009}).

\bibitem[{\citenamefont{Chen et~al.}(2010)\citenamefont{Chen, Pereira, and
  Balents}}]{CPB10}
\bibinfo{author}{\bibfnamefont{G.}~\bibnamefont{Chen}},
  \bibinfo{author}{\bibfnamefont{R.}~\bibnamefont{Pereira}}, \bibnamefont{and}
  \bibinfo{author}{\bibfnamefont{L.}~\bibnamefont{Balents}},
  \bibinfo{journal}{Phys. Rev. B} \textbf{\bibinfo{volume}{82}},
  \bibinfo{pages}{174440} (\bibinfo{year}{2010}).

\bibitem[{\citenamefont{Witczak-Krempa
  et~al.}(2014)\citenamefont{Witczak-Krempa, Chen, Kim, and Balents}}]{WCK+14}
\bibinfo{author}{\bibfnamefont{W.}~\bibnamefont{Witczak-Krempa}},
  \bibinfo{author}{\bibfnamefont{G.}~\bibnamefont{Chen}},
  \bibinfo{author}{\bibfnamefont{Y.~B.} \bibnamefont{Kim}}, \bibnamefont{and}
  \bibinfo{author}{\bibfnamefont{L.}~\bibnamefont{Balents}},
  \bibinfo{journal}{Annu. Rev. Condens. Matter Phys.}
  \textbf{\bibinfo{volume}{5}}, \bibinfo{pages}{57} (\bibinfo{year}{2014}).

\bibitem[{\citenamefont{Kim et~al.}(2008)\citenamefont{Kim, Jin, Moon, Kim,
  Park, Leem, Yu, Noh, Kim, Oh et~al.}}]{KJM+08}
\bibinfo{author}{\bibfnamefont{B.~J.} \bibnamefont{Kim}},
  \bibinfo{author}{\bibfnamefont{H.}~\bibnamefont{Jin}},
  \bibinfo{author}{\bibfnamefont{S.~J.} \bibnamefont{Moon}},
  \bibinfo{author}{\bibfnamefont{J.-Y.} \bibnamefont{Kim}},
  \bibinfo{author}{\bibfnamefont{B.-G.} \bibnamefont{Park}},
  \bibinfo{author}{\bibfnamefont{C.~S.} \bibnamefont{Leem}},
  \bibinfo{author}{\bibfnamefont{J.}~\bibnamefont{Yu}},
  \bibinfo{author}{\bibfnamefont{T.~W.} \bibnamefont{Noh}},
  \bibinfo{author}{\bibfnamefont{C.}~\bibnamefont{Kim}},
  \bibinfo{author}{\bibfnamefont{S.-J.} \bibnamefont{Oh}},
  \bibnamefont{et~al.}, \bibinfo{journal}{Phys. Rev. Lett.}
  \textbf{\bibinfo{volume}{101}}, \bibinfo{pages}{076402}
  (\bibinfo{year}{2008}).

\bibitem[{\citenamefont{Martins et~al.}(2011)\citenamefont{Martins, Aichhorn,
  Vaugier, and Biermann}}]{MAV+11}
\bibinfo{author}{\bibfnamefont{C.}~\bibnamefont{Martins}},
  \bibinfo{author}{\bibfnamefont{M.}~\bibnamefont{Aichhorn}},
  \bibinfo{author}{\bibfnamefont{L.}~\bibnamefont{Vaugier}}, \bibnamefont{and}
  \bibinfo{author}{\bibfnamefont{S.}~\bibnamefont{Biermann}},
  \bibinfo{journal}{Phys. Rev. Lett.} \textbf{\bibinfo{volume}{107}},
  \bibinfo{pages}{266404} (\bibinfo{year}{2011}).

\bibitem[{\citenamefont{Antonov et~al.}(2018)\citenamefont{Antonov, Uba, and
  Uba}}]{AUU18}
\bibinfo{author}{\bibfnamefont{V.~N.} \bibnamefont{Antonov}},
  \bibinfo{author}{\bibfnamefont{S.}~\bibnamefont{Uba}}, \bibnamefont{and}
  \bibinfo{author}{\bibfnamefont{L.}~\bibnamefont{Uba}},
  \bibinfo{journal}{Phys. Rev. B} \textbf{\bibinfo{volume}{98}},
  \bibinfo{pages}{245113} (\bibinfo{year}{2018}).

\bibitem[{\citenamefont{Qi and Zhang}(2010)}]{QZ10}
\bibinfo{author}{\bibfnamefont{X.-L.} \bibnamefont{Qi}} \bibnamefont{and}
  \bibinfo{author}{\bibfnamefont{S.-C.} \bibnamefont{Zhang}},
  \bibinfo{journal}{Physics Today} \textbf{\bibinfo{volume}{63}},
  \bibinfo{pages}{33} (\bibinfo{year}{2010}).

\bibitem[{\citenamefont{Ando}(2013)}]{Ando13}
\bibinfo{author}{\bibfnamefont{Y.}~\bibnamefont{Ando}}, \bibinfo{journal}{J.
  Phys. Soc. Jpn.} \textbf{\bibinfo{volume}{82}}, \bibinfo{pages}{102001}
  (\bibinfo{year}{2013}).

\bibitem[{\citenamefont{Wehling et~al.}(2014)\citenamefont{Wehling,
  Black-Schafferc, and Balatsky}}]{WBB14}
\bibinfo{author}{\bibfnamefont{T.~O.} \bibnamefont{Wehling}},
  \bibinfo{author}{\bibfnamefont{A.}~\bibnamefont{Black-Schafferc}},
  \bibnamefont{and} \bibinfo{author}{\bibfnamefont{A.}~\bibnamefont{Balatsky}},
  \bibinfo{journal}{Adv. Phys.} \textbf{\bibinfo{volume}{63}},
  \bibinfo{pages}{1} (\bibinfo{year}{2014}).

\bibitem[{\citenamefont{Bansi et~al.}(2016)\citenamefont{Bansi, H, and
  Das}}]{BLD16}
\bibinfo{author}{\bibfnamefont{A.}~\bibnamefont{Bansi}},
  \bibinfo{author}{\bibfnamefont{L.}~\bibnamefont{H}}, \bibnamefont{and}
  \bibinfo{author}{\bibfnamefont{T.}~\bibnamefont{Das}}, \bibinfo{journal}{Rev.
  Mod. Phys.} \textbf{\bibinfo{volume}{88}}, \bibinfo{pages}{021004}
  (\bibinfo{year}{2016}).

\bibitem[{\citenamefont{Kim et~al.}(2009)\citenamefont{Kim, Ohsumi, Komesu,
  Sakai, Morita, Takagi, and Arima}}]{KOK+09}
\bibinfo{author}{\bibfnamefont{B.~J.} \bibnamefont{Kim}},
  \bibinfo{author}{\bibfnamefont{H.}~\bibnamefont{Ohsumi}},
  \bibinfo{author}{\bibfnamefont{T.}~\bibnamefont{Komesu}},
  \bibinfo{author}{\bibfnamefont{S.}~\bibnamefont{Sakai}},
  \bibinfo{author}{\bibfnamefont{T.}~\bibnamefont{Morita}},
  \bibinfo{author}{\bibfnamefont{H.}~\bibnamefont{Takagi}}, \bibnamefont{and}
  \bibinfo{author}{\bibfnamefont{T.}~\bibnamefont{Arima}},
  \bibinfo{journal}{Science} \textbf{\bibinfo{volume}{323}},
  \bibinfo{pages}{1329} (\bibinfo{year}{2009}).

\bibitem[{\citenamefont{Watanabe et~al.}(2010)\citenamefont{Watanabe,
  Shirakawa, and Yunoki}}]{WSY10}
\bibinfo{author}{\bibfnamefont{H.}~\bibnamefont{Watanabe}},
  \bibinfo{author}{\bibfnamefont{T.}~\bibnamefont{Shirakawa}},
  \bibnamefont{and} \bibinfo{author}{\bibfnamefont{S.}~\bibnamefont{Yunoki}},
  \bibinfo{journal}{Phys. Rev. Lett.} \textbf{\bibinfo{volume}{105}},
  \bibinfo{pages}{216410} (\bibinfo{year}{2010}).

\bibitem[{\citenamefont{Witczak-Krempa and Kim}(2012)}]{WiKi12}
\bibinfo{author}{\bibfnamefont{W.}~\bibnamefont{Witczak-Krempa}}
  \bibnamefont{and} \bibinfo{author}{\bibfnamefont{Y.~B.} \bibnamefont{Kim}},
  \bibinfo{journal}{Phys. Rev. B} \textbf{\bibinfo{volume}{85}},
  \bibinfo{pages}{045124} (\bibinfo{year}{2012}).

\bibitem[{\citenamefont{Go et~al.}(2012)\citenamefont{Go, Witczak-Krempa, Jeon,
  Park, and Kim}}]{GWJ12}
\bibinfo{author}{\bibfnamefont{A.}~\bibnamefont{Go}},
  \bibinfo{author}{\bibfnamefont{W.}~\bibnamefont{Witczak-Krempa}},
  \bibinfo{author}{\bibfnamefont{G.~S.} \bibnamefont{Jeon}},
  \bibinfo{author}{\bibfnamefont{K.}~\bibnamefont{Park}}, \bibnamefont{and}
  \bibinfo{author}{\bibfnamefont{Y.~B.} \bibnamefont{Kim}},
  \bibinfo{journal}{Phys. Rev. Lett.} \textbf{\bibinfo{volume}{109}},
  \bibinfo{pages}{066401} (\bibinfo{year}{2012}).

\bibitem[{\citenamefont{Sushkov et~al.}(2015)\citenamefont{Sushkov, Hofmann,
  Jenkins, Ishikawa, Nakatsuji, DasSarma, and Drew}}]{SHJ+15}
\bibinfo{author}{\bibfnamefont{A.~B.} \bibnamefont{Sushkov}},
  \bibinfo{author}{\bibfnamefont{J.~B.} \bibnamefont{Hofmann}},
  \bibinfo{author}{\bibfnamefont{G.~S.} \bibnamefont{Jenkins}},
  \bibinfo{author}{\bibfnamefont{J.}~\bibnamefont{Ishikawa}},
  \bibinfo{author}{\bibfnamefont{S.}~\bibnamefont{Nakatsuji}},
  \bibinfo{author}{\bibfnamefont{S.}~\bibnamefont{DasSarma}}, \bibnamefont{and}
  \bibinfo{author}{\bibfnamefont{H.~D.} \bibnamefont{Drew}},
  \bibinfo{journal}{Phys. Rev. B} \textbf{\bibinfo{volume}{92}},
  \bibinfo{pages}{241108} (\bibinfo{year}{2015}).

\bibitem[{\citenamefont{Kimchi et~al.}(2014)\citenamefont{Kimchi, Analytis, and
  Vishwanath}}]{KAV14}
\bibinfo{author}{\bibfnamefont{I.}~\bibnamefont{Kimchi}},
  \bibinfo{author}{\bibfnamefont{J.~G.} \bibnamefont{Analytis}},
  \bibnamefont{and}
  \bibinfo{author}{\bibfnamefont{A.}~\bibnamefont{Vishwanath}},
  \bibinfo{journal}{Phys. Rev. B} \textbf{\bibinfo{volume}{90}},
  \bibinfo{pages}{205126} (\bibinfo{year}{2014}).

\bibitem[{\citenamefont{Kim et~al.}(2015)\citenamefont{Kim, Jin, Moon, Kim,
  Park, Leem, Yu, Noh, Kim, Park et~al.}}]{KJM+15}
\bibinfo{author}{\bibfnamefont{B.~J.} \bibnamefont{Kim}},
  \bibinfo{author}{\bibfnamefont{H.}~\bibnamefont{Jin}},
  \bibinfo{author}{\bibfnamefont{S.~J.} \bibnamefont{Moon}},
  \bibinfo{author}{\bibfnamefont{J.-Y.} \bibnamefont{Kim}},
  \bibinfo{author}{\bibfnamefont{B.-G.} \bibnamefont{Park}},
  \bibinfo{author}{\bibfnamefont{C.~S.} \bibnamefont{Leem}},
  \bibinfo{author}{\bibfnamefont{J.}~\bibnamefont{Yu}},
  \bibinfo{author}{\bibfnamefont{T.~W.} \bibnamefont{Noh}},
  \bibinfo{author}{\bibfnamefont{C.}~\bibnamefont{Kim}},
  \bibinfo{author}{\bibfnamefont{S.-J.-H.} \bibnamefont{Park}},
  \bibnamefont{et~al.}, \bibinfo{journal}{Phys. Rev. B}
  \textbf{\bibinfo{volume}{115}}, \bibinfo{pages}{176402}
  (\bibinfo{year}{2015}).

\bibitem[{\citenamefont{Yan et~al.}(2015)\citenamefont{Yan, Ren, Xu, Xie, Tao,
  Choi, Lee, Choi, Zhang, and Feng}}]{YRX+15}
\bibinfo{author}{\bibfnamefont{Y.}~\bibnamefont{Yan}},
  \bibinfo{author}{\bibfnamefont{M.}~\bibnamefont{Ren}},
  \bibinfo{author}{\bibfnamefont{H.}~\bibnamefont{Xu}},
  \bibinfo{author}{\bibfnamefont{B.}~\bibnamefont{Xie}},
  \bibinfo{author}{\bibfnamefont{R.}~\bibnamefont{Tao}},
  \bibinfo{author}{\bibfnamefont{H.}~\bibnamefont{Choi}},
  \bibinfo{author}{\bibfnamefont{N.}~\bibnamefont{Lee}},
  \bibinfo{author}{\bibfnamefont{Y.}~\bibnamefont{Choi}},
  \bibinfo{author}{\bibfnamefont{T.}~\bibnamefont{Zhang}}, \bibnamefont{and}
  \bibinfo{author}{\bibfnamefont{D.}~\bibnamefont{Feng}},
  \bibinfo{journal}{Phys. Rev. X} \textbf{\bibinfo{volume}{5}},
  \bibinfo{pages}{041018} (\bibinfo{year}{2015}).

\bibitem[{\citenamefont{Battisti et~al.}(2017)\citenamefont{Battisti,
  Bastiaans, Fedoseev, de~la Torre, Iliopoulos, Tamai, Hunter, Perry, Zaanen,
  Baumberger et~al.}}]{BBF+17}
\bibinfo{author}{\bibfnamefont{I.}~\bibnamefont{Battisti}},
  \bibinfo{author}{\bibfnamefont{K.~M.} \bibnamefont{Bastiaans}},
  \bibinfo{author}{\bibfnamefont{V.}~\bibnamefont{Fedoseev}},
  \bibinfo{author}{\bibfnamefont{A.}~\bibnamefont{de~la Torre}},
  \bibinfo{author}{\bibfnamefont{N.}~\bibnamefont{Iliopoulos}},
  \bibinfo{author}{\bibfnamefont{A.}~\bibnamefont{Tamai}},
  \bibinfo{author}{\bibfnamefont{E.~C.} \bibnamefont{Hunter}},
  \bibinfo{author}{\bibfnamefont{R.~S.} \bibnamefont{Perry}},
  \bibinfo{author}{\bibfnamefont{J.}~\bibnamefont{Zaanen}},
  \bibinfo{author}{\bibfnamefont{F.}~\bibnamefont{Baumberger}},
  \bibnamefont{et~al.}, \bibinfo{journal}{Nat. Phys.}
  \textbf{\bibinfo{volume}{13}}, \bibinfo{pages}{21} (\bibinfo{year}{2017}).

\bibitem[{\citenamefont{Damascelli et~al.}(2003)\citenamefont{Damascelli,
  Hussain, and Shen}}]{DHS03}
\bibinfo{author}{\bibfnamefont{A.}~\bibnamefont{Damascelli}},
  \bibinfo{author}{\bibfnamefont{Z.}~\bibnamefont{Hussain}}, \bibnamefont{and}
  \bibinfo{author}{\bibfnamefont{Z.-X.} \bibnamefont{Shen}},
  \bibinfo{journal}{Rev. Mod. Phys.} \textbf{\bibinfo{volume}{75}},
  \bibinfo{pages}{473} (\bibinfo{year}{2003}).

\bibitem[{\citenamefont{Kim et~al.}(2014)\citenamefont{Kim, Krupin, Denlinger,
  Bostwick, Rotenberg, Zhao, Mitchell, Allen, and Kim}}]{KKD+14}
\bibinfo{author}{\bibfnamefont{Y.~K.} \bibnamefont{Kim}},
  \bibinfo{author}{\bibfnamefont{O.}~\bibnamefont{Krupin}},
  \bibinfo{author}{\bibfnamefont{J.~D.} \bibnamefont{Denlinger}},
  \bibinfo{author}{\bibfnamefont{A.}~\bibnamefont{Bostwick}},
  \bibinfo{author}{\bibfnamefont{E.}~\bibnamefont{Rotenberg}},
  \bibinfo{author}{\bibfnamefont{Q.}~\bibnamefont{Zhao}},
  \bibinfo{author}{\bibfnamefont{J.~F.} \bibnamefont{Mitchell}},
  \bibinfo{author}{\bibfnamefont{J.~W.} \bibnamefont{Allen}}, \bibnamefont{and}
  \bibinfo{author}{\bibfnamefont{B.~J.} \bibnamefont{Kim}},
  \bibinfo{journal}{Science} \textbf{\bibinfo{volume}{345}},
  \bibinfo{pages}{187} (\bibinfo{year}{2014}).

\bibitem[{\citenamefont{Kim et~al.}(2016)\citenamefont{Kim, Sung, Denlinger,
  and Kim}}]{KSD+16}
\bibinfo{author}{\bibfnamefont{Y.~K.} \bibnamefont{Kim}},
  \bibinfo{author}{\bibfnamefont{N.~H.} \bibnamefont{Sung}},
  \bibinfo{author}{\bibfnamefont{J.~D.} \bibnamefont{Denlinger}},
  \bibnamefont{and} \bibinfo{author}{\bibfnamefont{B.~J.} \bibnamefont{Kim}},
  \bibinfo{journal}{Nat. Phys.} \textbf{\bibinfo{volume}{12}},
  \bibinfo{pages}{37} (\bibinfo{year}{2016}).

\bibitem[{\citenamefont{Terashima et~al.}(2017)\citenamefont{Terashima,
  Sunagawa, Fujiwara, Fukura, Fujii, Okada, Horigane, Kobayashi, Horie,
  Akimitsu et~al.}}]{TSF+17}
\bibinfo{author}{\bibfnamefont{K.}~\bibnamefont{Terashima}},
  \bibinfo{author}{\bibfnamefont{M.}~\bibnamefont{Sunagawa}},
  \bibinfo{author}{\bibfnamefont{H.}~\bibnamefont{Fujiwara}},
  \bibinfo{author}{\bibfnamefont{T.}~\bibnamefont{Fukura}},
  \bibinfo{author}{\bibfnamefont{M.}~\bibnamefont{Fujii}},
  \bibinfo{author}{\bibfnamefont{K.}~\bibnamefont{Okada}},
  \bibinfo{author}{\bibfnamefont{K.}~\bibnamefont{Horigane}},
  \bibinfo{author}{\bibfnamefont{K.}~\bibnamefont{Kobayashi}},
  \bibinfo{author}{\bibfnamefont{R.}~\bibnamefont{Horie}},
  \bibinfo{author}{\bibfnamefont{J.}~\bibnamefont{Akimitsu}},
  \bibnamefont{et~al.}, \bibinfo{journal}{Phys. Rev. B}
  \textbf{\bibinfo{volume}{96}}, \bibinfo{pages}{041106(R)}
  (\bibinfo{year}{2017}).

\bibitem[{\citenamefont{Hu et~al.}(2019)\citenamefont{Hu, Chen, Peng, Lane,
  Matzelle, Sun, Hashimoto, Lu, Schwier, Arita et~al.}}]{HCP+19}
\bibinfo{author}{\bibfnamefont{Y.}~\bibnamefont{Hu}},
  \bibinfo{author}{\bibfnamefont{X.}~\bibnamefont{Chen}},
  \bibinfo{author}{\bibfnamefont{S.-T.} \bibnamefont{Peng}},
  \bibinfo{author}{\bibfnamefont{C.}~\bibnamefont{Lane}},
  \bibinfo{author}{\bibfnamefont{M.}~\bibnamefont{Matzelle}},
  \bibinfo{author}{\bibfnamefont{Z.-L.} \bibnamefont{Sun}},
  \bibinfo{author}{\bibfnamefont{M.}~\bibnamefont{Hashimoto}},
  \bibinfo{author}{\bibfnamefont{D.-H.} \bibnamefont{Lu}},
  \bibinfo{author}{\bibfnamefont{E.}~\bibnamefont{Schwier}},
  \bibinfo{author}{\bibfnamefont{M.}~\bibnamefont{Arita}},
  \bibnamefont{et~al.}, \bibinfo{journal}{Phys. Rev. Lett.}
  \textbf{\bibinfo{volume}{123}}, \bibinfo{pages}{216402}
  (\bibinfo{year}{2019}).

\bibitem[{\citenamefont{Kim et~al.}(2012{\natexlab{a}})\citenamefont{Kim, Casa,
  Upton, Gog, Kim, Mitchell, van Veenendaal, Daghofer, van~den Brink,
  Khaliullin et~al.}}]{KCU+12}
\bibinfo{author}{\bibfnamefont{J.}~\bibnamefont{Kim}},
  \bibinfo{author}{\bibfnamefont{D.}~\bibnamefont{Casa}},
  \bibinfo{author}{\bibfnamefont{M.~H.} \bibnamefont{Upton}},
  \bibinfo{author}{\bibfnamefont{T.}~\bibnamefont{Gog}},
  \bibinfo{author}{\bibfnamefont{Y.-J.} \bibnamefont{Kim}},
  \bibinfo{author}{\bibfnamefont{J.~F.} \bibnamefont{Mitchell}},
  \bibinfo{author}{\bibfnamefont{M.}~\bibnamefont{van Veenendaal}},
  \bibinfo{author}{\bibfnamefont{M.}~\bibnamefont{Daghofer}},
  \bibinfo{author}{\bibfnamefont{J.}~\bibnamefont{van~den Brink}},
  \bibinfo{author}{\bibfnamefont{G.}~\bibnamefont{Khaliullin}},
  \bibnamefont{et~al.}, \bibinfo{journal}{Phys. Rev. Lett.}
  \textbf{\bibinfo{volume}{108}}, \bibinfo{pages}{177003}
  (\bibinfo{year}{2012}{\natexlab{a}}).

\bibitem[{\citenamefont{Clancy et~al.}(2019)\citenamefont{Clancy, Gretarsson,
  Upton, Kim, Cao, and Kim}}]{YCY+19}
\bibinfo{author}{\bibfnamefont{J.~P.} \bibnamefont{Clancy}},
  \bibinfo{author}{\bibfnamefont{H.}~\bibnamefont{Gretarsson}},
  \bibinfo{author}{\bibfnamefont{M.~H.} \bibnamefont{Upton}},
  \bibinfo{author}{\bibfnamefont{J.}~\bibnamefont{Kim}},
  \bibinfo{author}{\bibfnamefont{G.}~\bibnamefont{Cao}}, \bibnamefont{and}
  \bibinfo{author}{\bibfnamefont{Y.-J.} \bibnamefont{Kim}},
  \bibinfo{journal}{Phys. Rev. B} \textbf{\bibinfo{volume}{100}},
  \bibinfo{pages}{104414} (\bibinfo{year}{2019}).

\bibitem[{\citenamefont{Bertinshaw et~al.}(2019)\citenamefont{Bertinshaw, Kim,
  Khaliullin, and Kim}}]{BKK19}
\bibinfo{author}{\bibfnamefont{J.}~\bibnamefont{Bertinshaw}},
  \bibinfo{author}{\bibfnamefont{Y.}~\bibnamefont{Kim}},
  \bibinfo{author}{\bibfnamefont{G.}~\bibnamefont{Khaliullin}},
  \bibnamefont{and} \bibinfo{author}{\bibfnamefont{B.}~\bibnamefont{Kim}},
  \bibinfo{journal}{Annu. Rev. Condens. Matter Phys.}
  \textbf{\bibinfo{volume}{7}}, \bibinfo{pages}{315} (\bibinfo{year}{2019}).

\bibitem[{\citenamefont{Kao et~al.}(1996)\citenamefont{Kao, Caliebe, Hastings,
  and Gillet}}]{KCH+96}
\bibinfo{author}{\bibfnamefont{C.-C.} \bibnamefont{Kao}},
  \bibinfo{author}{\bibfnamefont{W.~A.~L.} \bibnamefont{Caliebe}},
  \bibinfo{author}{\bibfnamefont{J.~B.} \bibnamefont{Hastings}},
  \bibnamefont{and} \bibinfo{author}{\bibfnamefont{J.-M.}
  \bibnamefont{Gillet}}, \bibinfo{journal}{Phys. Rev. B}
  \textbf{\bibinfo{volume}{54}}, \bibinfo{pages}{16361} (\bibinfo{year}{1996}).

\bibitem[{\citenamefont{Ament et~al.}(2011)\citenamefont{Ament, van Veenendaal,
  Devereaux, Hill, and van~den Brink}}]{AVD+11}
\bibinfo{author}{\bibfnamefont{L.~J.~P.} \bibnamefont{Ament}},
  \bibinfo{author}{\bibfnamefont{M.}~\bibnamefont{van Veenendaal}},
  \bibinfo{author}{\bibfnamefont{T.~P.} \bibnamefont{Devereaux}},
  \bibinfo{author}{\bibfnamefont{J.~P.} \bibnamefont{Hill}}, \bibnamefont{and}
  \bibinfo{author}{\bibfnamefont{J.}~\bibnamefont{van~den Brink}},
  \bibinfo{journal}{Rev. Mod. Phys.} \textbf{\bibinfo{volume}{83}},
  \bibinfo{pages}{705} (\bibinfo{year}{2011}).

\bibitem[{\citenamefont{Liu et~al.}(2012)\citenamefont{Liu, Katukuri, Hozoi,
  Yin, Dean, Upton, Kim, Casa, Said, Gog et~al.}}]{LKH+12}
\bibinfo{author}{\bibfnamefont{X.}~\bibnamefont{Liu}},
  \bibinfo{author}{\bibfnamefont{V.~M.} \bibnamefont{Katukuri}},
  \bibinfo{author}{\bibfnamefont{L.}~\bibnamefont{Hozoi}},
  \bibinfo{author}{\bibfnamefont{W.-G.} \bibnamefont{Yin}},
  \bibinfo{author}{\bibfnamefont{M.~P.~M.} \bibnamefont{Dean}},
  \bibinfo{author}{\bibfnamefont{M.~H.} \bibnamefont{Upton}},
  \bibinfo{author}{\bibfnamefont{J.}~\bibnamefont{Kim}},
  \bibinfo{author}{\bibfnamefont{D.}~\bibnamefont{Casa}},
  \bibinfo{author}{\bibfnamefont{A.}~\bibnamefont{Said}},
  \bibinfo{author}{\bibfnamefont{T.}~\bibnamefont{Gog}}, \bibnamefont{et~al.},
  \bibinfo{journal}{Phys. Rev. Lett.} \textbf{\bibinfo{volume}{109}},
  \bibinfo{pages}{157401} (\bibinfo{year}{2012}).

\bibitem[{\citenamefont{Hozoi et~al.}(2014)\citenamefont{Hozoi, Gretarsson,
  Clancy, Jeon, Lee, Kim, Yushankhai, Fulde, Casa, Gog et~al.}}]{HGC+14}
\bibinfo{author}{\bibfnamefont{L.}~\bibnamefont{Hozoi}},
  \bibinfo{author}{\bibfnamefont{H.}~\bibnamefont{Gretarsson}},
  \bibinfo{author}{\bibfnamefont{J.~P.} \bibnamefont{Clancy}},
  \bibinfo{author}{\bibfnamefont{B.-G.} \bibnamefont{Jeon}},
  \bibinfo{author}{\bibfnamefont{B.}~\bibnamefont{Lee}},
  \bibinfo{author}{\bibfnamefont{K.~H.} \bibnamefont{Kim}},
  \bibinfo{author}{\bibfnamefont{V.}~\bibnamefont{Yushankhai}},
  \bibinfo{author}{\bibfnamefont{P.}~\bibnamefont{Fulde}},
  \bibinfo{author}{\bibfnamefont{D.}~\bibnamefont{Casa}},
  \bibinfo{author}{\bibfnamefont{T.}~\bibnamefont{Gog}}, \bibnamefont{et~al.},
  \bibinfo{journal}{Phys. Rev. B} \textbf{\bibinfo{volume}{89}},
  \bibinfo{pages}{115111} (\bibinfo{year}{2014}).

\bibitem[{\citenamefont{Clancy et~al.}(2016)\citenamefont{Clancy, Gretarsson,
  Lee, Tian, Kim, Upton, Casa, Gog, Islam, Jeon et~al.}}]{CGL+16}
\bibinfo{author}{\bibfnamefont{J.~P.} \bibnamefont{Clancy}},
  \bibinfo{author}{\bibfnamefont{H.}~\bibnamefont{Gretarsson}},
  \bibinfo{author}{\bibfnamefont{E.~K.~H.} \bibnamefont{Lee}},
  \bibinfo{author}{\bibfnamefont{D.}~\bibnamefont{Tian}},
  \bibinfo{author}{\bibfnamefont{J.}~\bibnamefont{Kim}},
  \bibinfo{author}{\bibfnamefont{M.~H.} \bibnamefont{Upton}},
  \bibinfo{author}{\bibfnamefont{D.}~\bibnamefont{Casa}},
  \bibinfo{author}{\bibfnamefont{T.}~\bibnamefont{Gog}},
  \bibinfo{author}{\bibfnamefont{Z.}~\bibnamefont{Islam}},
  \bibinfo{author}{\bibfnamefont{B.-G.} \bibnamefont{Jeon}},
  \bibnamefont{et~al.}, \bibinfo{journal}{Phys. Rev. B}
  \textbf{\bibinfo{volume}{94}}, \bibinfo{pages}{024408}
  (\bibinfo{year}{2016}).

\bibitem[{\citenamefont{Nag et~al.}(2018)\citenamefont{Nag, Bhowal,
  Chakraborty, Sala, Efimenko, Bert, Biswas, Hillier, Itoh, Kaushik
  et~al.}}]{NBC+18}
\bibinfo{author}{\bibfnamefont{A.}~\bibnamefont{Nag}},
  \bibinfo{author}{\bibfnamefont{S.}~\bibnamefont{Bhowal}},
  \bibinfo{author}{\bibfnamefont{A.}~\bibnamefont{Chakraborty}},
  \bibinfo{author}{\bibfnamefont{M.~M.} \bibnamefont{Sala}},
  \bibinfo{author}{\bibfnamefont{A.}~\bibnamefont{Efimenko}},
  \bibinfo{author}{\bibfnamefont{F.}~\bibnamefont{Bert}},
  \bibinfo{author}{\bibfnamefont{P.~K.} \bibnamefont{Biswas}},
  \bibinfo{author}{\bibfnamefont{A.~D.} \bibnamefont{Hillier}},
  \bibinfo{author}{\bibfnamefont{M.}~\bibnamefont{Itoh}},
  \bibinfo{author}{\bibfnamefont{S.~D.} \bibnamefont{Kaushik}},
  \bibnamefont{et~al.}, \bibinfo{journal}{Phys. Rev. B}
  \textbf{\bibinfo{volume}{98}}, \bibinfo{pages}{014431}
  (\bibinfo{year}{2018}).

\bibitem[{\citenamefont{Takayama et~al.}(2019)\citenamefont{Takayama,
  Krajewska, Gibbs, Yaresko, Ishii, Yamaoka, Ishii, Hiraoka, Funnell, Bull
  et~al.}}]{TKG+19}
\bibinfo{author}{\bibfnamefont{T.}~\bibnamefont{Takayama}},
  \bibinfo{author}{\bibfnamefont{A.}~\bibnamefont{Krajewska}},
  \bibinfo{author}{\bibfnamefont{A.~S.} \bibnamefont{Gibbs}},
  \bibinfo{author}{\bibfnamefont{A.~N.} \bibnamefont{Yaresko}},
  \bibinfo{author}{\bibfnamefont{H.}~\bibnamefont{Ishii}},
  \bibinfo{author}{\bibfnamefont{H.}~\bibnamefont{Yamaoka}},
  \bibinfo{author}{\bibfnamefont{K.}~\bibnamefont{Ishii}},
  \bibinfo{author}{\bibfnamefont{N.}~\bibnamefont{Hiraoka}},
  \bibinfo{author}{\bibfnamefont{N.~P.} \bibnamefont{Funnell}},
  \bibinfo{author}{\bibfnamefont{C.~L.} \bibnamefont{Bull}},
  \bibnamefont{et~al.}, \bibinfo{journal}{Phys. Rev. B}
  \textbf{\bibinfo{volume}{99}}, \bibinfo{pages}{125127}
  (\bibinfo{year}{2019}).

\bibitem[{\citenamefont{Aczel et~al.}(2022)\citenamefont{Aczel, Chen, Clancy,
  dela Cruz, i~Plessis, MacDougall, Pollock, Upton, Williams, LaManna
  et~al.}}]{ACC+22}
\bibinfo{author}{\bibfnamefont{A.~A.} \bibnamefont{Aczel}},
  \bibinfo{author}{\bibfnamefont{Q.}~\bibnamefont{Chen}},
  \bibinfo{author}{\bibfnamefont{J.~P.} \bibnamefont{Clancy}},
  \bibinfo{author}{\bibfnamefont{C.}~\bibnamefont{dela Cruz}},
  \bibinfo{author}{\bibfnamefont{D.~R.} \bibnamefont{i~Plessis}},
  \bibinfo{author}{\bibfnamefont{G.~J.} \bibnamefont{MacDougall}},
  \bibinfo{author}{\bibfnamefont{C.~J.} \bibnamefont{Pollock}},
  \bibinfo{author}{\bibfnamefont{M.~H.} \bibnamefont{Upton}},
  \bibinfo{author}{\bibfnamefont{T.~J.} \bibnamefont{Williams}},
  \bibinfo{author}{\bibfnamefont{N.}~\bibnamefont{LaManna}},
  \bibnamefont{et~al.}, \bibinfo{journal}{Phys. Rev. Mater.}
  \textbf{\bibinfo{volume}{6}}, \bibinfo{pages}{094409} (\bibinfo{year}{2022}).

\bibitem[{\citenamefont{Calder et~al.}(2016)\citenamefont{Calder, Vale,
  Bogdanov, Liu, Donnerer, Upton, Casa, Said, Lumsden, Zhao et~al.}}]{CVB+16}
\bibinfo{author}{\bibfnamefont{S.}~\bibnamefont{Calder}},
  \bibinfo{author}{\bibfnamefont{J.}~\bibnamefont{Vale}},
  \bibinfo{author}{\bibfnamefont{N.}~\bibnamefont{Bogdanov}},
  \bibinfo{author}{\bibfnamefont{X.}~\bibnamefont{Liu}},
  \bibinfo{author}{\bibfnamefont{C.}~\bibnamefont{Donnerer}},
  \bibinfo{author}{\bibfnamefont{M.}~\bibnamefont{Upton}},
  \bibinfo{author}{\bibfnamefont{D.}~\bibnamefont{Casa}},
  \bibinfo{author}{\bibfnamefont{A.}~\bibnamefont{Said}},
  \bibinfo{author}{\bibfnamefont{M.}~\bibnamefont{Lumsden}},
  \bibinfo{author}{\bibfnamefont{Z.}~\bibnamefont{Zhao}}, \bibnamefont{et~al.},
  \bibinfo{journal}{Nature Commun.} \textbf{\bibinfo{volume}{7}},
  \bibinfo{pages}{11651} (\bibinfo{year}{2016}).

\bibitem[{\citenamefont{Taylor et~al.}(2017)\citenamefont{Taylor, Calder,
  Morrow, Feng, Upton, Lumsden, Yamaura, Woodward, and Christianson}}]{TCM+17}
\bibinfo{author}{\bibfnamefont{A.~E.} \bibnamefont{Taylor}},
  \bibinfo{author}{\bibfnamefont{S.}~\bibnamefont{Calder}},
  \bibinfo{author}{\bibfnamefont{R.}~\bibnamefont{Morrow}},
  \bibinfo{author}{\bibfnamefont{H.~L.} \bibnamefont{Feng}},
  \bibinfo{author}{\bibfnamefont{M.~H.} \bibnamefont{Upton}},
  \bibinfo{author}{\bibfnamefont{M.~D.} \bibnamefont{Lumsden}},
  \bibinfo{author}{\bibfnamefont{K.}~\bibnamefont{Yamaura}},
  \bibinfo{author}{\bibfnamefont{P.~M.} \bibnamefont{Woodward}},
  \bibnamefont{and} \bibinfo{author}{\bibfnamefont{A.~D.}
  \bibnamefont{Christianson}}, \bibinfo{journal}{Phys. Rev. Lett.}
  \textbf{\bibinfo{volume}{118}}, \bibinfo{pages}{207202}
  (\bibinfo{year}{2017}).

\bibitem[{\citenamefont{Calder et~al.}(2017)\citenamefont{Calder, Vale,
  Bogdanov, Donnerer, Pincini, Sala, Liu, Upton, Casa, Shi et~al.}}]{CVB+17}
\bibinfo{author}{\bibfnamefont{S.}~\bibnamefont{Calder}},
  \bibinfo{author}{\bibfnamefont{J.~G.} \bibnamefont{Vale}},
  \bibinfo{author}{\bibfnamefont{N.}~\bibnamefont{Bogdanov}},
  \bibinfo{author}{\bibfnamefont{C.}~\bibnamefont{Donnerer}},
  \bibinfo{author}{\bibfnamefont{D.}~\bibnamefont{Pincini}},
  \bibinfo{author}{\bibfnamefont{M.~M.} \bibnamefont{Sala}},
  \bibinfo{author}{\bibfnamefont{X.}~\bibnamefont{Liu}},
  \bibinfo{author}{\bibfnamefont{M.~H.} \bibnamefont{Upton}},
  \bibinfo{author}{\bibfnamefont{D.}~\bibnamefont{Casa}},
  \bibinfo{author}{\bibfnamefont{Y.~G.} \bibnamefont{Shi}},
  \bibnamefont{et~al.}, \bibinfo{journal}{Phys. Rev. B}
  \textbf{\bibinfo{volume}{95}}, \bibinfo{pages}{020413(R)}
  (\bibinfo{year}{2017}).

\bibitem[{\citenamefont{Liu et~al.}(2015)\citenamefont{Liu, Dean, Liu,
  Chiuzbaian, Jaouen, Nicolaou, Yin, Serrao, Ramesh, Ding et~al.}}]{LDL+15}
\bibinfo{author}{\bibfnamefont{X.}~\bibnamefont{Liu}},
  \bibinfo{author}{\bibfnamefont{M.~P.~M.} \bibnamefont{Dean}},
  \bibinfo{author}{\bibfnamefont{J.}~\bibnamefont{Liu}},
  \bibinfo{author}{\bibfnamefont{S.~G.} \bibnamefont{Chiuzbaian}},
  \bibinfo{author}{\bibfnamefont{N.}~\bibnamefont{Jaouen}},
  \bibinfo{author}{\bibfnamefont{A.}~\bibnamefont{Nicolaou}},
  \bibinfo{author}{\bibfnamefont{W.~G.} \bibnamefont{Yin}},
  \bibinfo{author}{\bibfnamefont{C.~R.} \bibnamefont{Serrao}},
  \bibinfo{author}{\bibfnamefont{R.}~\bibnamefont{Ramesh}},
  \bibinfo{author}{\bibfnamefont{H.}~\bibnamefont{Ding}}, \bibnamefont{et~al.},
  \bibinfo{journal}{J. Phys.: Condens. Matter} \textbf{\bibinfo{volume}{27}},
  \bibinfo{pages}{202202} (\bibinfo{year}{2015}).

\bibitem[{\citenamefont{Monney et~al.}(2020)\citenamefont{Monney, Herzog,
  Pulkkinen, Huang, Pelliciari, Olalde-Velasco, Katayama, Nohara, Takagi,
  Schmitt et~al.}}]{MHP+20}
\bibinfo{author}{\bibfnamefont{C.}~\bibnamefont{Monney}},
  \bibinfo{author}{\bibfnamefont{M.}~\bibnamefont{Herzog}},
  \bibinfo{author}{\bibfnamefont{A.}~\bibnamefont{Pulkkinen}},
  \bibinfo{author}{\bibfnamefont{Y.}~\bibnamefont{Huang}},
  \bibinfo{author}{\bibfnamefont{J.}~\bibnamefont{Pelliciari}},
  \bibinfo{author}{\bibfnamefont{P.}~\bibnamefont{Olalde-Velasco}},
  \bibinfo{author}{\bibfnamefont{N.}~\bibnamefont{Katayama}},
  \bibinfo{author}{\bibfnamefont{M.}~\bibnamefont{Nohara}},
  \bibinfo{author}{\bibfnamefont{H.}~\bibnamefont{Takagi}},
  \bibinfo{author}{\bibfnamefont{T.}~\bibnamefont{Schmitt}},
  \bibnamefont{et~al.}, \bibinfo{journal}{Phys. Rev. B}
  \textbf{\bibinfo{volume}{102}}, \bibinfo{pages}{085148}
  (\bibinfo{year}{2020}).

\bibitem[{\citenamefont{Lee et~al.}(2013)\citenamefont{Lee, Johnston, Moritz,
  Lee, Yi, Zhou, Schmitt, Patthey, Strocov, Kudo et~al.}}]{LJM+13}
\bibinfo{author}{\bibfnamefont{W.~S.} \bibnamefont{Lee}},
  \bibinfo{author}{\bibfnamefont{S.}~\bibnamefont{Johnston}},
  \bibinfo{author}{\bibfnamefont{B.}~\bibnamefont{Moritz}},
  \bibinfo{author}{\bibfnamefont{J.}~\bibnamefont{Lee}},
  \bibinfo{author}{\bibfnamefont{M.}~\bibnamefont{Yi}},
  \bibinfo{author}{\bibfnamefont{K.~J.} \bibnamefont{Zhou}},
  \bibinfo{author}{\bibfnamefont{T.}~\bibnamefont{Schmitt}},
  \bibinfo{author}{\bibfnamefont{L.}~\bibnamefont{Patthey}},
  \bibinfo{author}{\bibfnamefont{V.}~\bibnamefont{Strocov}},
  \bibinfo{author}{\bibfnamefont{K.}~\bibnamefont{Kudo}}, \bibnamefont{et~al.},
  \bibinfo{journal}{Phys. Rev. Lett.} \textbf{\bibinfo{volume}{110}},
  \bibinfo{pages}{265502} (\bibinfo{year}{2013}).

\bibitem[{\citenamefont{Lu et~al.}(2018)\citenamefont{Lu, Olalde-Velasco,
  Huang, Bisogni, Pelliciari, Fatale, Dantz, Vale, Hunter, Chang
  et~al.}}]{LOH+18}
\bibinfo{author}{\bibfnamefont{X.}~\bibnamefont{Lu}},
  \bibinfo{author}{\bibfnamefont{P.}~\bibnamefont{Olalde-Velasco}},
  \bibinfo{author}{\bibfnamefont{Y.}~\bibnamefont{Huang}},
  \bibinfo{author}{\bibfnamefont{V.}~\bibnamefont{Bisogni}},
  \bibinfo{author}{\bibfnamefont{J.}~\bibnamefont{Pelliciari}},
  \bibinfo{author}{\bibfnamefont{S.}~\bibnamefont{Fatale}},
  \bibinfo{author}{\bibfnamefont{M.}~\bibnamefont{Dantz}},
  \bibinfo{author}{\bibfnamefont{J.~G.} \bibnamefont{Vale}},
  \bibinfo{author}{\bibfnamefont{E.~C.} \bibnamefont{Hunter}},
  \bibinfo{author}{\bibfnamefont{J.}~\bibnamefont{Chang}},
  \bibnamefont{et~al.}, \bibinfo{journal}{Phys. Rev. B}
  \textbf{\bibinfo{volume}{97}}, \bibinfo{pages}{041102(R)}
  (\bibinfo{year}{2018}).

\bibitem[{\citenamefont{Kim et~al.}(2023)\citenamefont{Kim, Dietl, Kim, Ha,
  Kim, Said, Kim, and Kim}}]{KDK+23}
\bibinfo{author}{\bibfnamefont{J.-K.} \bibnamefont{Kim}},
  \bibinfo{author}{\bibfnamefont{C.}~\bibnamefont{Dietl}},
  \bibinfo{author}{\bibfnamefont{H.-W.~J.} \bibnamefont{Kim}},
  \bibinfo{author}{\bibfnamefont{S.-H.} \bibnamefont{Ha}},
  \bibinfo{author}{\bibfnamefont{J.}~\bibnamefont{Kim}},
  \bibinfo{author}{\bibfnamefont{A.~H.} \bibnamefont{Said}},
  \bibinfo{author}{\bibfnamefont{J.}~\bibnamefont{Kim}}, \bibnamefont{and}
  \bibinfo{author}{\bibfnamefont{B.~J.} \bibnamefont{Kim}},
  \bibinfo{journal}{J. Synchrotron Rad.} \textbf{\bibinfo{volume}{30}},
  \bibinfo{pages}{643} (\bibinfo{year}{2023}).

\bibitem[{\citenamefont{Paris et~al.}(2020)\citenamefont{Paris, Tseng,
  P\"arschke, Zhang, Upton, Efimenko, Rolfs, , McNally, Maurel
  et~al.}}]{PTP+20}
\bibinfo{author}{\bibfnamefont{E.}~\bibnamefont{Paris}},
  \bibinfo{author}{\bibfnamefont{Y.}~\bibnamefont{Tseng}},
  \bibinfo{author}{\bibfnamefont{E.~M.} \bibnamefont{P\"arschke}},
  \bibinfo{author}{\bibfnamefont{W.}~\bibnamefont{Zhang}},
  \bibinfo{author}{\bibfnamefont{M.~H.} \bibnamefont{Upton}},
  \bibinfo{author}{\bibfnamefont{A.}~\bibnamefont{Efimenko}},
  \bibinfo{author}{\bibfnamefont{K.}~\bibnamefont{Rolfs}}, ,
  \bibinfo{author}{\bibfnamefont{D.~E.} \bibnamefont{McNally}},
  \bibinfo{author}{\bibfnamefont{L.}~\bibnamefont{Maurel}},
  \bibnamefont{et~al.}, \bibinfo{journal}{Proceed. Nat. Acad. Sciences USA}
  \textbf{\bibinfo{volume}{117}}, \bibinfo{pages}{24764}
  (\bibinfo{year}{2020}).

\bibitem[{\citenamefont{Ishii et~al.}(2011)\citenamefont{Ishii, Jarrige,
  Yoshida, Ikeuchi, Mizuki, Ohashi, Takayama, Matsuno, and Takagi}}]{IJY+11}
\bibinfo{author}{\bibfnamefont{K.}~\bibnamefont{Ishii}},
  \bibinfo{author}{\bibfnamefont{I.}~\bibnamefont{Jarrige}},
  \bibinfo{author}{\bibfnamefont{M.}~\bibnamefont{Yoshida}},
  \bibinfo{author}{\bibfnamefont{K.}~\bibnamefont{Ikeuchi}},
  \bibinfo{author}{\bibfnamefont{J.}~\bibnamefont{Mizuki}},
  \bibinfo{author}{\bibfnamefont{K.}~\bibnamefont{Ohashi}},
  \bibinfo{author}{\bibfnamefont{T.}~\bibnamefont{Takayama}},
  \bibinfo{author}{\bibfnamefont{J.}~\bibnamefont{Matsuno}}, \bibnamefont{and}
  \bibinfo{author}{\bibfnamefont{H.}~\bibnamefont{Takagi}},
  \bibinfo{journal}{Phys. Rev. B} \textbf{\bibinfo{volume}{83}},
  \bibinfo{pages}{115121} (\bibinfo{year}{2011}).

\bibitem[{\citenamefont{Lupascu et~al.}(2014)\citenamefont{Lupascu, Clancy,
  Gretarsson, Nie, Nichols, Terzic, Cao, Seo, Islam, and et~al.}}]{LCG+14}
\bibinfo{author}{\bibfnamefont{A.}~\bibnamefont{Lupascu}},
  \bibinfo{author}{\bibfnamefont{J.~P.} \bibnamefont{Clancy}},
  \bibinfo{author}{\bibfnamefont{H.}~\bibnamefont{Gretarsson}},
  \bibinfo{author}{\bibfnamefont{Z.}~\bibnamefont{Nie}},
  \bibinfo{author}{\bibfnamefont{J.}~\bibnamefont{Nichols}},
  \bibinfo{author}{\bibfnamefont{J.}~\bibnamefont{Terzic}},
  \bibinfo{author}{\bibfnamefont{G.}~\bibnamefont{Cao}},
  \bibinfo{author}{\bibfnamefont{S.~S.~A.} \bibnamefont{Seo}},
  \bibinfo{author}{\bibfnamefont{Z.}~\bibnamefont{Islam}}, \bibnamefont{and}
  \bibinfo{author}{\bibfnamefont{M.~H.~U.} \bibnamefont{et~al.}},
  \bibinfo{journal}{Phys. Rev. Lett.} \textbf{\bibinfo{volume}{112}},
  \bibinfo{pages}{147201} (\bibinfo{year}{2014}).

\bibitem[{\citenamefont{Bertinshaw et~al.}(2020)\citenamefont{Bertinshaw, Kim,
  Porras, Ueda, Sung, Efimenko, Bombardi, Kim, Keimer, and Kim}}]{BKP+20}
\bibinfo{author}{\bibfnamefont{J.}~\bibnamefont{Bertinshaw}},
  \bibinfo{author}{\bibfnamefont{J.~K.} \bibnamefont{Kim}},
  \bibinfo{author}{\bibfnamefont{J.}~\bibnamefont{Porras}},
  \bibinfo{author}{\bibfnamefont{K.}~\bibnamefont{Ueda}},
  \bibinfo{author}{\bibfnamefont{N.~H.} \bibnamefont{Sung}},
  \bibinfo{author}{\bibfnamefont{A.}~\bibnamefont{Efimenko}},
  \bibinfo{author}{\bibfnamefont{A.}~\bibnamefont{Bombardi}},
  \bibinfo{author}{\bibfnamefont{J.}~\bibnamefont{Kim}},
  \bibinfo{author}{\bibfnamefont{B.}~\bibnamefont{Keimer}}, \bibnamefont{and}
  \bibinfo{author}{\bibfnamefont{B.~J.} \bibnamefont{Kim}},
  \bibinfo{journal}{Phys. Rev. B} \textbf{\bibinfo{volume}{101}},
  \bibinfo{pages}{094428} (\bibinfo{year}{2020}).

\bibitem[{\citenamefont{Clancy et~al.}(2023)\citenamefont{Clancy, Gretarsson,
  Lupascu, Sears, Nie, Upton, Kim, Islam, Uchida, Schlom et~al.}}]{CGL+23}
\bibinfo{author}{\bibfnamefont{J.~P.} \bibnamefont{Clancy}},
  \bibinfo{author}{\bibfnamefont{H.}~\bibnamefont{Gretarsson}},
  \bibinfo{author}{\bibfnamefont{A.}~\bibnamefont{Lupascu}},
  \bibinfo{author}{\bibfnamefont{J.~A.} \bibnamefont{Sears}},
  \bibinfo{author}{\bibfnamefont{Z.}~\bibnamefont{Nie}},
  \bibinfo{author}{\bibfnamefont{M.~H.} \bibnamefont{Upton}},
  \bibinfo{author}{\bibfnamefont{J.}~\bibnamefont{Kim}},
  \bibinfo{author}{\bibfnamefont{Z.}~\bibnamefont{Islam}},
  \bibinfo{author}{\bibfnamefont{M.}~\bibnamefont{Uchida}},
  \bibinfo{author}{\bibfnamefont{D.~G.} \bibnamefont{Schlom}},
  \bibnamefont{et~al.}, \bibinfo{journal}{Phys. Rev. B}
  \textbf{\bibinfo{volume}{107}}, \bibinfo{pages}{054423}
  (\bibinfo{year}{2023}).

\bibitem[{\citenamefont{Guo et~al.}(1994)\citenamefont{Guo, Ebert, Temmerman,
  and Durham}}]{GET+94}
\bibinfo{author}{\bibfnamefont{G.~Y.} \bibnamefont{Guo}},
  \bibinfo{author}{\bibfnamefont{H.}~\bibnamefont{Ebert}},
  \bibinfo{author}{\bibfnamefont{W.~M.} \bibnamefont{Temmerman}},
  \bibnamefont{and} \bibinfo{author}{\bibfnamefont{P.~J.}
  \bibnamefont{Durham}}, \bibinfo{journal}{Phys. Rev. B}
  \textbf{\bibinfo{volume}{50}}, \bibinfo{pages}{3861} (\bibinfo{year}{1994}).

\bibitem[{\citenamefont{Antonov et~al.}(2004)\citenamefont{Antonov, Harmon, and
  Yaresko}}]{book:AHY04}
\bibinfo{author}{\bibfnamefont{V.}~\bibnamefont{Antonov}},
  \bibinfo{author}{\bibfnamefont{B.}~\bibnamefont{Harmon}}, \bibnamefont{and}
  \bibinfo{author}{\bibfnamefont{A.}~\bibnamefont{Yaresko}},
  \emph{\bibinfo{title}{Electronic Structure and Magneto-Optical Properties of
  Solids}} (\bibinfo{publisher}{Kluwer}, \bibinfo{address}{Dordrecht},
  \bibinfo{year}{2004}).

\bibitem[{\citenamefont{Arola et~al.}(2004)\citenamefont{Arola, Horne, Strange,
  Winter, Szotek, and Temmerman}}]{AHS+04}
\bibinfo{author}{\bibfnamefont{E.}~\bibnamefont{Arola}},
  \bibinfo{author}{\bibfnamefont{M.}~\bibnamefont{Horne}},
  \bibinfo{author}{\bibfnamefont{P.}~\bibnamefont{Strange}},
  \bibinfo{author}{\bibfnamefont{H.}~\bibnamefont{Winter}},
  \bibinfo{author}{\bibfnamefont{Z.}~\bibnamefont{Szotek}}, \bibnamefont{and}
  \bibinfo{author}{\bibfnamefont{W.~M.} \bibnamefont{Temmerman}},
  \bibinfo{journal}{Phys. Rev. B} \textbf{\bibinfo{volume}{70}},
  \bibinfo{pages}{235127} (\bibinfo{year}{2004}).

\bibitem[{\citenamefont{Nemoshkalenko et~al.}(1983)\citenamefont{Nemoshkalenko,
  Krasovskii, Antonov, Antonov, Fleck, Wonn, and Ziesche}}]{NKA+83}
\bibinfo{author}{\bibfnamefont{V.~V.} \bibnamefont{Nemoshkalenko}},
  \bibinfo{author}{\bibfnamefont{A.~E.} \bibnamefont{Krasovskii}},
  \bibinfo{author}{\bibfnamefont{V.~N.} \bibnamefont{Antonov}},
  \bibinfo{author}{\bibfnamefont{V.~N.} \bibnamefont{Antonov}},
  \bibinfo{author}{\bibfnamefont{U.}~\bibnamefont{Fleck}},
  \bibinfo{author}{\bibfnamefont{H.}~\bibnamefont{Wonn}}, \bibnamefont{and}
  \bibinfo{author}{\bibfnamefont{P.}~\bibnamefont{Ziesche}},
  \bibinfo{journal}{Phys. status solidi B} \textbf{\bibinfo{volume}{120}},
  \bibinfo{pages}{283} (\bibinfo{year}{1983}).

\bibitem[{\citenamefont{Arola et~al.}(1997{\natexlab{a}})\citenamefont{Arola,
  Strange, and Gyorffy}}]{ASG97}
\bibinfo{author}{\bibfnamefont{E.}~\bibnamefont{Arola}},
  \bibinfo{author}{\bibfnamefont{P.}~\bibnamefont{Strange}}, \bibnamefont{and}
  \bibinfo{author}{\bibfnamefont{B.~L.} \bibnamefont{Gyorffy}},
  \bibinfo{journal}{Phys. Rev. B} \textbf{\bibinfo{volume}{55}},
  \bibinfo{pages}{472} (\bibinfo{year}{1997}{\natexlab{a}}).

\bibitem[{\citenamefont{Arola et~al.}(1997{\natexlab{b}})\citenamefont{Arola,
  Strange, and Gyorffy}}]{ASG97E}
\bibinfo{author}{\bibfnamefont{E.}~\bibnamefont{Arola}},
  \bibinfo{author}{\bibfnamefont{P.}~\bibnamefont{Strange}}, \bibnamefont{and}
  \bibinfo{author}{\bibfnamefont{B.~L.} \bibnamefont{Gyorffy}},
  \bibinfo{journal}{Phys. Rev. B} \textbf{\bibinfo{volume}{55}},
  \bibinfo{pages}{472} (\bibinfo{year}{1997}{\natexlab{b}}).

\bibitem[{\citenamefont{Antonov
  et~al.}(2022{\natexlab{a}})\citenamefont{Antonov, Kukusta, and
  Bekenov}}]{AKB22a}
\bibinfo{author}{\bibfnamefont{V.~N.} \bibnamefont{Antonov}},
  \bibinfo{author}{\bibfnamefont{D.~A.} \bibnamefont{Kukusta}},
  \bibnamefont{and} \bibinfo{author}{\bibfnamefont{L.~V.}
  \bibnamefont{Bekenov}}, \bibinfo{journal}{Phys. Rev. B}
  \textbf{\bibinfo{volume}{105}}, \bibinfo{pages}{155144}
  (\bibinfo{year}{2022}{\natexlab{a}}).

\bibitem[{\citenamefont{Shimura et~al.}(1995)\citenamefont{Shimura, Inaguma,
  Nakamura, Itoh, and Morii}}]{SIN+95}
\bibinfo{author}{\bibfnamefont{T.}~\bibnamefont{Shimura}},
  \bibinfo{author}{\bibfnamefont{Y.}~\bibnamefont{Inaguma}},
  \bibinfo{author}{\bibfnamefont{T.}~\bibnamefont{Nakamura}},
  \bibinfo{author}{\bibfnamefont{M.}~\bibnamefont{Itoh}}, \bibnamefont{and}
  \bibinfo{author}{\bibfnamefont{Y.}~\bibnamefont{Morii}},
  \bibinfo{journal}{Phys. Rev. B} \textbf{\bibinfo{volume}{52}},
  \bibinfo{pages}{9143} (\bibinfo{year}{1995}).

\bibitem[{\citenamefont{Antonov et~al.}(2006)\citenamefont{Antonov, Jepsen,
  Yaresko, and Shpak}}]{AJY+06}
\bibinfo{author}{\bibfnamefont{V.~N.} \bibnamefont{Antonov}},
  \bibinfo{author}{\bibfnamefont{O.}~\bibnamefont{Jepsen}},
  \bibinfo{author}{\bibfnamefont{A.~N.} \bibnamefont{Yaresko}},
  \bibnamefont{and} \bibinfo{author}{\bibfnamefont{A.~P.} \bibnamefont{Shpak}},
  \bibinfo{journal}{J. Appl. Phys.} \textbf{\bibinfo{volume}{100}},
  \bibinfo{pages}{043711} (\bibinfo{year}{2006}).

\bibitem[{\citenamefont{Antonov et~al.}(2007)\citenamefont{Antonov, Harmon,
  Yaresko, and Shpak}}]{AHY+07b}
\bibinfo{author}{\bibfnamefont{V.~N.} \bibnamefont{Antonov}},
  \bibinfo{author}{\bibfnamefont{B.~N.} \bibnamefont{Harmon}},
  \bibinfo{author}{\bibfnamefont{A.~N.} \bibnamefont{Yaresko}},
  \bibnamefont{and} \bibinfo{author}{\bibfnamefont{A.~P.} \bibnamefont{Shpak}},
  \bibinfo{journal}{Phys. Rev. B} \textbf{\bibinfo{volume}{75}},
  \bibinfo{pages}{184422} (\bibinfo{year}{2007}).

\bibitem[{\citenamefont{Antonov et~al.}(2010)\citenamefont{Antonov, Yaresko,
  and Jepsen}}]{AYJ10}
\bibinfo{author}{\bibfnamefont{V.~N.} \bibnamefont{Antonov}},
  \bibinfo{author}{\bibfnamefont{A.~N.} \bibnamefont{Yaresko}},
  \bibnamefont{and} \bibinfo{author}{\bibfnamefont{O.}~\bibnamefont{Jepsen}},
  \bibinfo{journal}{Phys. Rev. B} \textbf{\bibinfo{volume}{81}},
  \bibinfo{pages}{075209} (\bibinfo{year}{2010}).

\bibitem[{\citenamefont{Andersen}(1975)}]{And75}
\bibinfo{author}{\bibfnamefont{O.~K.} \bibnamefont{Andersen}},
  \bibinfo{journal}{Phys. Rev. B} \textbf{\bibinfo{volume}{12}},
  \bibinfo{pages}{3060} (\bibinfo{year}{1975}).

\bibitem[{\citenamefont{Perdew et~al.}(1996)\citenamefont{Perdew, Burke, and
  Ernzerhof}}]{PBE96}
\bibinfo{author}{\bibfnamefont{J.~P.} \bibnamefont{Perdew}},
  \bibinfo{author}{\bibfnamefont{K.}~\bibnamefont{Burke}}, \bibnamefont{and}
  \bibinfo{author}{\bibfnamefont{M.}~\bibnamefont{Ernzerhof}},
  \bibinfo{journal}{Phys. Rev. Lett.} \textbf{\bibinfo{volume}{77}},
  \bibinfo{pages}{3865} (\bibinfo{year}{1996}).

\bibitem[{\citenamefont{Bl\"ochl et~al.}(1994)\citenamefont{Bl\"ochl, Jepsen,
  and Andersen}}]{BJA94}
\bibinfo{author}{\bibfnamefont{P.~E.} \bibnamefont{Bl\"ochl}},
  \bibinfo{author}{\bibfnamefont{O.}~\bibnamefont{Jepsen}}, \bibnamefont{and}
  \bibinfo{author}{\bibfnamefont{O.~K.} \bibnamefont{Andersen}},
  \bibinfo{journal}{Phys. Rev. B} \textbf{\bibinfo{volume}{49}},
  \bibinfo{pages}{16223} (\bibinfo{year}{1994}).

\bibitem[{\citenamefont{Yaresko et~al.}(2003)\citenamefont{Yaresko, Antonov,
  and Fulde}}]{YAF03}
\bibinfo{author}{\bibfnamefont{A.~N.} \bibnamefont{Yaresko}},
  \bibinfo{author}{\bibfnamefont{V.~N.} \bibnamefont{Antonov}},
  \bibnamefont{and} \bibinfo{author}{\bibfnamefont{P.}~\bibnamefont{Fulde}},
  \bibinfo{journal}{Phys. Rev. B} \textbf{\bibinfo{volume}{67}},
  \bibinfo{pages}{155103} (\bibinfo{year}{2003}).

\bibitem[{\citenamefont{Dederichs et~al.}(1984)\citenamefont{Dederichs,
  Bl\"ugel, Zeller, and Akai}}]{DBZ+84}
\bibinfo{author}{\bibfnamefont{P.~H.} \bibnamefont{Dederichs}},
  \bibinfo{author}{\bibfnamefont{S.}~\bibnamefont{Bl\"ugel}},
  \bibinfo{author}{\bibfnamefont{R.}~\bibnamefont{Zeller}}, \bibnamefont{and}
  \bibinfo{author}{\bibfnamefont{H.}~\bibnamefont{Akai}},
  \bibinfo{journal}{Phys. Rev. Lett.} \textbf{\bibinfo{volume}{53}},
  \bibinfo{pages}{2512} (\bibinfo{year}{1984}).

\bibitem[{\citenamefont{Pickett et~al.}(1998)\citenamefont{Pickett, Erwin, and
  Ethridge}}]{PEE98}
\bibinfo{author}{\bibfnamefont{W.~E.} \bibnamefont{Pickett}},
  \bibinfo{author}{\bibfnamefont{S.~C.} \bibnamefont{Erwin}}, \bibnamefont{and}
  \bibinfo{author}{\bibfnamefont{E.~C.} \bibnamefont{Ethridge}},
  \bibinfo{journal}{Phys. Rev. B} \textbf{\bibinfo{volume}{58}},
  \bibinfo{pages}{1201} (\bibinfo{year}{1998}).

\bibitem[{\citenamefont{Antonov
  et~al.}(2022{\natexlab{b}})\citenamefont{Antonov, Kukusta, and
  Bekenov}}]{AKB22b}
\bibinfo{author}{\bibfnamefont{V.~N.} \bibnamefont{Antonov}},
  \bibinfo{author}{\bibfnamefont{D.~A.} \bibnamefont{Kukusta}},
  \bibnamefont{and} \bibinfo{author}{\bibfnamefont{L.~V.}
  \bibnamefont{Bekenov}}, \bibinfo{journal}{Phys. Rev. B}
  \textbf{\bibinfo{volume}{105}}, \bibinfo{pages}{155145}
  (\bibinfo{year}{2022}{\natexlab{b}}).

\bibitem[{\citenamefont{Pr\"opper et~al.}(2016)\citenamefont{Pr\"opper,
  Yaresko, H\"oppner, Matiks, Mathis, Takayama, Matsumoto, Takagi, Keimer, and
  Boris}}]{PYH+16a}
\bibinfo{author}{\bibfnamefont{D.}~\bibnamefont{Pr\"opper}},
  \bibinfo{author}{\bibfnamefont{A.~N.} \bibnamefont{Yaresko}},
  \bibinfo{author}{\bibfnamefont{M.}~\bibnamefont{H\"oppner}},
  \bibinfo{author}{\bibfnamefont{Y.}~\bibnamefont{Matiks}},
  \bibinfo{author}{\bibfnamefont{Y.-L.} \bibnamefont{Mathis}},
  \bibinfo{author}{\bibfnamefont{T.}~\bibnamefont{Takayama}},
  \bibinfo{author}{\bibfnamefont{A.}~\bibnamefont{Matsumoto}},
  \bibinfo{author}{\bibfnamefont{H.}~\bibnamefont{Takagi}},
  \bibinfo{author}{\bibfnamefont{B.}~\bibnamefont{Keimer}}, \bibnamefont{and}
  \bibinfo{author}{\bibfnamefont{A.~V.} \bibnamefont{Boris}},
  \bibinfo{journal}{Phys. Rev. B} \textbf{\bibinfo{volume}{94}},
  \bibinfo{pages}{035158} (\bibinfo{year}{2016}).

\bibitem[{\citenamefont{Zhang et~al.}(2013)\citenamefont{Zhang, Haule, and
  Vanderbilt}}]{ZHV13}
\bibinfo{author}{\bibfnamefont{H.}~\bibnamefont{Zhang}},
  \bibinfo{author}{\bibfnamefont{K.}~\bibnamefont{Haule}}, \bibnamefont{and}
  \bibinfo{author}{\bibfnamefont{D.}~\bibnamefont{Vanderbilt}},
  \bibinfo{journal}{Phys. Rev. Lett.} \textbf{\bibinfo{volume}{111}},
  \bibinfo{pages}{246402} (\bibinfo{year}{2013}).

\bibitem[{\citenamefont{Pr\"opper et~al.}(2013)\citenamefont{Pr\"opper,
  Yaresko, H\"oppner, Matiks, Mathis, Takayama, Matsumoto, Takagi, Keimer,
  Boris et~al.}}]{PYH+16}
\bibinfo{author}{\bibfnamefont{D.}~\bibnamefont{Pr\"opper}},
  \bibinfo{author}{\bibfnamefont{A.~N.} \bibnamefont{Yaresko}},
  \bibinfo{author}{\bibfnamefont{M.}~\bibnamefont{H\"oppner}},
  \bibinfo{author}{\bibfnamefont{Y.}~\bibnamefont{Matiks}},
  \bibinfo{author}{\bibfnamefont{Y.-L.} \bibnamefont{Mathis}},
  \bibinfo{author}{\bibfnamefont{T.}~\bibnamefont{Takayama}},
  \bibinfo{author}{\bibfnamefont{A.}~\bibnamefont{Matsumoto}},
  \bibinfo{author}{\bibfnamefont{H.}~\bibnamefont{Takagi}},
  \bibinfo{author}{\bibfnamefont{B.}~\bibnamefont{Keimer}},
  \bibinfo{author}{\bibfnamefont{A.~V.} \bibnamefont{Boris}},
  \bibnamefont{et~al.}, \bibinfo{journal}{Phys. Rev. B}
  \textbf{\bibinfo{volume}{88}}, \bibinfo{pages}{085125}
  (\bibinfo{year}{2013}).

\bibitem[{\citenamefont{Wang et~al.}(2013)\citenamefont{Wang, Cao, Waugh, Park,
  Qi, Korneta, Cao, and Dessau}}]{WCW+13}
\bibinfo{author}{\bibfnamefont{Q.}~\bibnamefont{Wang}},
  \bibinfo{author}{\bibfnamefont{Y.}~\bibnamefont{Cao}},
  \bibinfo{author}{\bibfnamefont{J.~A.} \bibnamefont{Waugh}},
  \bibinfo{author}{\bibfnamefont{S.~R.} \bibnamefont{Park}},
  \bibinfo{author}{\bibfnamefont{T.~F.} \bibnamefont{Qi}},
  \bibinfo{author}{\bibfnamefont{O.~B.} \bibnamefont{Korneta}},
  \bibinfo{author}{\bibfnamefont{G.}~\bibnamefont{Cao}}, \bibnamefont{and}
  \bibinfo{author}{\bibfnamefont{D.~S.} \bibnamefont{Dessau}},
  \bibinfo{journal}{Phys. Rev. B} \textbf{\bibinfo{volume}{87}},
  \bibinfo{pages}{245109} (\bibinfo{year}{2013}).

\bibitem[{\citenamefont{Kim et~al.}(2012{\natexlab{b}})\citenamefont{Kim,
  Khaliullin, and Min}}]{KKM12}
\bibinfo{author}{\bibfnamefont{B.~H.} \bibnamefont{Kim}},
  \bibinfo{author}{\bibfnamefont{G.}~\bibnamefont{Khaliullin}},
  \bibnamefont{and} \bibinfo{author}{\bibfnamefont{B.~I.} \bibnamefont{Min}},
  \bibinfo{journal}{Phys. Rev. Lett.} \textbf{\bibinfo{volume}{109}},
  \bibinfo{pages}{167205} (\bibinfo{year}{2012}{\natexlab{b}}).

\bibitem[{\citenamefont{Haskel et~al.}(2012)\citenamefont{Haskel, Fabbris,
  Zhernenkov, Kong, Jin, Cao, and van Veenendaal}}]{HFZ+12}
\bibinfo{author}{\bibfnamefont{D.}~\bibnamefont{Haskel}},
  \bibinfo{author}{\bibfnamefont{G.}~\bibnamefont{Fabbris}},
  \bibinfo{author}{\bibfnamefont{M.}~\bibnamefont{Zhernenkov}},
  \bibinfo{author}{\bibfnamefont{P.~P.} \bibnamefont{Kong}},
  \bibinfo{author}{\bibfnamefont{C.~Q.} \bibnamefont{Jin}},
  \bibinfo{author}{\bibfnamefont{G.}~\bibnamefont{Cao}}, \bibnamefont{and}
  \bibinfo{author}{\bibfnamefont{M.}~\bibnamefont{van Veenendaal}},
  \bibinfo{journal}{Phys. Rev. Lett.} \textbf{\bibinfo{volume}{109}},
  \bibinfo{pages}{027204} (\bibinfo{year}{2012}).

\bibitem[{\citenamefont{van~der Laan and Thole}(1988)}]{LaTh88}
\bibinfo{author}{\bibfnamefont{G.}~\bibnamefont{van~der Laan}}
  \bibnamefont{and} \bibinfo{author}{\bibfnamefont{B.~T.} \bibnamefont{Thole}},
  \bibinfo{journal}{Phys. Rev. Lett.} \textbf{\bibinfo{volume}{60}},
  \bibinfo{pages}{1977} (\bibinfo{year}{1988}).

\bibitem[{\citenamefont{Liu et~al.}(2008)\citenamefont{Liu, Antonov, Jepsen,
  and Andersen}}]{LAJ+08}
\bibinfo{author}{\bibfnamefont{G.-Q.} \bibnamefont{Liu}},
  \bibinfo{author}{\bibfnamefont{V.~N.} \bibnamefont{Antonov}},
  \bibinfo{author}{\bibfnamefont{O.}~\bibnamefont{Jepsen}}, \bibnamefont{and}
  \bibinfo{author}{\bibfnamefont{O.~K.} \bibnamefont{Andersen}},
  \bibinfo{journal}{Phys. Rev. Lett.} \textbf{\bibinfo{volume}{101}},
  \bibinfo{pages}{026408} (\bibinfo{year}{2008}).

\bibitem[{\citenamefont{Krajewska et~al.}(2020)\citenamefont{Krajewska,
  Takayama, Dinnebier, Yaresko, Ishii, Isobe, and Takagi}}]{KTD+20}
\bibinfo{author}{\bibfnamefont{A.}~\bibnamefont{Krajewska}},
  \bibinfo{author}{\bibfnamefont{T.}~\bibnamefont{Takayama}},
  \bibinfo{author}{\bibfnamefont{R.}~\bibnamefont{Dinnebier}},
  \bibinfo{author}{\bibfnamefont{A.}~\bibnamefont{Yaresko}},
  \bibinfo{author}{\bibfnamefont{K.}~\bibnamefont{Ishii}},
  \bibinfo{author}{\bibfnamefont{M.}~\bibnamefont{Isobe}}, \bibnamefont{and}
  \bibinfo{author}{\bibfnamefont{H.}~\bibnamefont{Takagi}},
  \bibinfo{journal}{Phys. Rev. B} \textbf{\bibinfo{volume}{101}},
  \bibinfo{pages}{121101(R)} (\bibinfo{year}{2020}).

\bibitem[{\citenamefont{Ishii et~al.}(2005)\citenamefont{Ishii, Tsutsui, Endoh,
  Tohyama, Maekawa, Hoesch, Kuzushita, Tsubota, Inami, Mizuki et~al.}}]{ITE+05}
\bibinfo{author}{\bibfnamefont{K.}~\bibnamefont{Ishii}},
  \bibinfo{author}{\bibfnamefont{K.}~\bibnamefont{Tsutsui}},
  \bibinfo{author}{\bibfnamefont{Y.}~\bibnamefont{Endoh}},
  \bibinfo{author}{\bibfnamefont{T.}~\bibnamefont{Tohyama}},
  \bibinfo{author}{\bibfnamefont{S.}~\bibnamefont{Maekawa}},
  \bibinfo{author}{\bibfnamefont{M.}~\bibnamefont{Hoesch}},
  \bibinfo{author}{\bibfnamefont{K.}~\bibnamefont{Kuzushita}},
  \bibinfo{author}{\bibfnamefont{M.}~\bibnamefont{Tsubota}},
  \bibinfo{author}{\bibfnamefont{T.}~\bibnamefont{Inami}},
  \bibinfo{author}{\bibfnamefont{J.}~\bibnamefont{Mizuki}},
  \bibnamefont{et~al.}, \bibinfo{journal}{Phys. Rev. Lett.}
  \textbf{\bibinfo{volume}{94}}, \bibinfo{pages}{207003}
  (\bibinfo{year}{2005}).

\bibitem[{\citenamefont{Kaneko et~al.}(2013)\citenamefont{Kaneko, Toriyama,
  Konishi, and Ohta}}]{KTK+13}
\bibinfo{author}{\bibfnamefont{T.}~\bibnamefont{Kaneko}},
  \bibinfo{author}{\bibfnamefont{T.}~\bibnamefont{Toriyama}},
  \bibinfo{author}{\bibfnamefont{T.}~\bibnamefont{Konishi}}, \bibnamefont{and}
  \bibinfo{author}{\bibfnamefont{Y.}~\bibnamefont{Ohta}},
  \bibinfo{journal}{Phys. Rev. B} \textbf{\bibinfo{volume}{87}},
  \bibinfo{pages}{035121} (\bibinfo{year}{2013}).

\bibitem[{\citenamefont{King et~al.}(2013)\citenamefont{King, Takayama, Tamai,
  Rozbicki, Walker, Shi, Patthey, Moore, Lu, Shen et~al.}}]{KTT+13}
\bibinfo{author}{\bibfnamefont{P.~D.~C.} \bibnamefont{King}},
  \bibinfo{author}{\bibfnamefont{T.}~\bibnamefont{Takayama}},
  \bibinfo{author}{\bibfnamefont{A.}~\bibnamefont{Tamai}},
  \bibinfo{author}{\bibfnamefont{E.}~\bibnamefont{Rozbicki}},
  \bibinfo{author}{\bibfnamefont{S.~M.} \bibnamefont{Walker}},
  \bibinfo{author}{\bibfnamefont{M.}~\bibnamefont{Shi}},
  \bibinfo{author}{\bibfnamefont{L.}~\bibnamefont{Patthey}},
  \bibinfo{author}{\bibfnamefont{R.~G.} \bibnamefont{Moore}},
  \bibinfo{author}{\bibfnamefont{D.}~\bibnamefont{Lu}},
  \bibinfo{author}{\bibfnamefont{K.~M.} \bibnamefont{Shen}},
  \bibnamefont{et~al.}, \bibinfo{journal}{Phys. Rev. B}
  \textbf{\bibinfo{volume}{87}}, \bibinfo{pages}{241106(R)}
  (\bibinfo{year}{2013}).

\end{thebibliography}

\newcommand{\noopsort}[1]{} \newcommand{\printfirst}[2]{#1}
  \newcommand{\singleletter}[1]{#1} \newcommand{\switchargs}[2]{#2#1}

\end{document}